\documentclass[aps,prl,preprint,superscriptaddress,showpacs]{revtex4-1}
\usepackage{epsfig}
\usepackage{graphicx}
\usepackage{dcolumn}
\usepackage{bm}
\usepackage{amsmath}
\usepackage{color}

\newcommand{\pop}{{\it Phys. Plasmas}}

\begin{document}
\title{
Enhanced toroidal flow stabilization of edge localized modes with increased plasma density
}

\author{Shikui Cheng}
\affiliation{CAS Key Laboratory of Geospace Environment and Department of Modern Physics, University of Science and Technology of China, Hefei, Anhui 230026, PRC}

\author{Ping Zhu}
\email[]{pzhu@ustc.edu.cn}
\affiliation{CAS Key Laboratory of Geospace Environment and Department of Modern Physics, University of Science and Technology of China, Hefei, Anhui 230026, PRC}
\affiliation{Department of Engineering Physics, University of Wisconsin-Madison, Madison, Wisconsin 53706, USA}

\author{Debabrata Banerjee}
\affiliation{CAS Key Laboratory of Geospace Environment and Department of Modern Physics, University of Science and Technology of China, Hefei, Anhui 230026, PRC}

\begin{abstract}
Toroidal flow alone is generally thought to have important influence on tokamak edge pedestal stability, even though theory analysis often predicts merely a weak stabilizing effect of toroidal flow on the edge localized modes (ELMs) in experimental parameter regimes. For the first time, we find from two-fluid MHD calculations that such a stabilization, however, can be significantly enhanced by increasing the edge plasma density. Our finding resolves a long-standing mystery whether or how toroidal rotation can indeed have effective influence on ELMs, and explains why the ELM mitigation and suppression by toroidal rotation are more favorably achieved in higher collisionality regime in recent experiments. The finding suggests a new control scheme on modulating toroidal flow stabilization of ELMs with plasma density, along with a new additional constraint on the optimal level of plasma density for the desired edge plasma conditions.
\end{abstract}



\maketitle
Plasma flows are widely believed to play significant roles in the dynamics, transport, and structural formation of natural and laboratory plasmas. Despite rapid progress in our understanding, the roles and effects of plasma flow continue to be a fundamental theme in vast different areas of plasma physics research. For example, in the astrophysical environment, plasma rotation and its instability are considered as a major mechanism for accretion disk momentum transport and jet formation~\citep{carey09a,Matthew16}. In Earth's magnetotail, observations of plasma flows suggests they are critical agents in triggering substorm onset and auroral brightening~\citep{angelopoulos92a,angelopoulos08c,nishimuray14b,nishimuray16a}. In laser plasma experiments, plasma flow is observed to drive magnetic reconnection and generate bubble-jet structure~\citep{Fox11,Plechaty13}. In magnetic fusion plasmas, plasma flow and rotation can effectively interact with both major MHD modes and microinstability-driven turbulence~\cite{Zohm96,Burrell16,terry00a}. One particular such interaction is that between the toroidal rotation and the edge localized modes (ELMs) in the high confinement mode (H-mode) of tokamaks~\cite{Wagner82,Zohm96,Burrell05,Burrell09}.

 

Type-I ELMs can carry up to $20\%$ of stored plasma energy outside of the confined domain and deposit high loads of heat and particles on plasma facing components~\citep{Leonard14,Loarte07}, which leads to considerable plasma confinement degradation and component material deterioration. To optimize fusion performance, it is imperative to simultaneously avoid ELMs while maximizing pedestal height. The NBI (neutral beam injection) induced toroidal rotation and its radial shear are believed to be responsible for transforming large-sized type-I ELMs to grassy ELMs with much smaller size in JT-60U~\citep{Oyama05}. With NBI heating, ELM-free quiescent H-mode (QH) has also been achieved in DIII-D~\citep{Burrell05,Burrell09}.



The effect of toroidal flow on edge localized modes has been a subject of long time studies in theory~\citep{Cooper88,Wealbroeck91,Furukawa05,Aiba09,Aiba10,Aiba11,Xi12,Xia13}. Whereas the sheared toroidal rotation alone has been found to be stabilizing on the high-$n$ edge localized modes, such effects are in general rather weak for modes in the range of intermediate-$n$ and for realistic flow magnitude in experimental regime. Here $n$ is the toroidal mode number. In fact, the toroidal shear flow is known to be even destabilizing on low-$n$ ELMs. Thus it remains a mystery whether or how toroidal rotation can indeed be effective in suppressing ELMs in tokamak experiments where the realistic Mach number of edge toroidal flow is rarely above $0.5$~\cite{mazzucato93a,rice96a,gormezano96a,conway00a,Vries06,Rice07,Politzer08}. The answer to this question may greatly impact the design and operation strategies for future burning plasma tokamaks such as ITER~\citep{Loarte07} and CFETR~\citep{Wan14,Chan15}.


On the other hand, plasma density is one of the key elements for governing the tokamak edge plasma conditions. Optimal level of plasma density in the edge region has been searched for in order to achieve the desired fusion gain and divertor heat/particle load mitigation. However, how density may affect the edge pedestal stability has not been well known. Recent experiments have found significant effects of density and hence collisionality on ELM instability. The amplitude of type-I ELM decreases with increasing edge collisionality in the presence of flow in JT-60U~\citep{Oyama10}. In DIII-D experiment, QH-modes are produced during NBI heating, which introduces both momentum and particle sources. As a consequence, plasma density increases from $0.5\times 10^{19}m^{-3}$ before NBI heating to $3\times 10^{19}m^{-3}$ after application of NBI in DIII-D shot\#136011, for example~\citep{Burrell09}. Similar results are also observed in EAST, where the reduction in ELM size due to sheared flow is larger for higher plasma density, and the mitigation and suppression of ELMs by toroidal rotation occur more favorably in regimes with higher collisionality~\citep{Kong}. In particular, the ELM size reduction due to rotation is $14\%$ with lower density ($3.56\times 10^{19}m^{-3}$), which is enhanced to $22.6\%$ at higher density ($4.13 \times 10^{19}m^{-3}$). The higher collisionality regime of edge pedestal region are mainly achieved from increasing the edge plasma density using gas-puffing. These experimental results suggest an exciting and new critical role of plasma density in directly influencing and modulating the stabilizing effects of toroidal flow on edge pedestal instabilities.

To verify and explore the potentially new roles of plasma density suggested above, we have carried out linear stability analysis using the initial-value extended full MHD code NIMROD~\cite{Sovinec04}, which has been benchmarked and verified for both ideal and non-ideal physical processes ~\cite{Sovinec10,Brennan06,Zhu13,Burke10,Ebrahimi15,Izzo15,Zhu08}. The extended MHD equations used in our NIMROD calculations are:

\begin{equation}
\frac{\partial n}{\partial t} + \nabla \cdot \left( n {\bf u}\right) = 0
\end{equation}
\begin{equation}
m n \left( \frac{\partial}{\partial t} + {\bf u} \cdot \nabla \right) {\bf u} = {\bf J} \times {\bf B} - \nabla p - \nabla \cdot \overline{\Pi}
\end{equation}
\begin{equation}
\frac{3}{2} \left( \frac{\partial}{\partial t} + {\bf u}_{\alpha} \cdot \nabla \right) T_{\alpha} = -n T_{\alpha} \nabla \cdot {\bf u}_{\alpha} - \nabla \cdot {\bf q}_{\alpha}\quad\quad (\alpha=i,e)
\end{equation}
\begin{equation}
\frac{\partial {\bf B}}{\partial t} = - \nabla \times \left[ \eta {\bf J} - {\bf u}\times{\bf B} + \frac{1}{ne} \left({\bf J}\times{\bf B} - \nabla p_e \right)\right]
\end{equation}
\begin{equation}
\mu_0 {\bf J} = \nabla \times {\bf B}, ~~~~~~~~~~~~~~~ \nabla \cdot {\bf B} = 0
\end{equation}
where {\bf u} is the center-of-mass flow velocity with particle density $n$ and ion mass $m$, $p$ is the combined pressure of electron ($p_e$) and ion ($p_i$), $\eta$ represents resistivity, ${\bf q}_{e,i}$
denote conductive heat flux vectors, $\overline{\Pi}$ is ion stress tensor including gyro-viscous components, and the rest of symbols are conventional as described in earlier references (e.g.~\citep{Sovinec10}).

For the first time, our study finds that increasing edge plasma density can indeed substantially enhance the stabilizing effect of toroidal flow shear on ELMs. Whereas increasing density alone is found destabilizing on ELM in absence of flow, the density increase can introduce enhanced stabilizing effects of toroidal flow on ELMs with even moderate Mach number. Our new finding provides the first theoretical explanation for the experimental observations of ELM stabilization by toroidal rotation in higher collisionality regimes in EAST~\citep{Kong}.

The rest of this paper is organized as follows. First, we report how density would affect the toroidal flow stabilizing effects on linear edge localized modes for a model tokamak equilibrium with edge pedestal within the single-fluid MHD model.  Second, we demonstrate that the density effects persist even in two-fluid MHD model. Finally, we summarize our findings and discuss how they may help us understand the recent EAST experiments on ELM mitigation and suppression.

\begin{figure}[ht]
\begin{minipage}{0.49\textwidth}
  \includegraphics[width=1.0\textwidth,height=0.3\textheight]{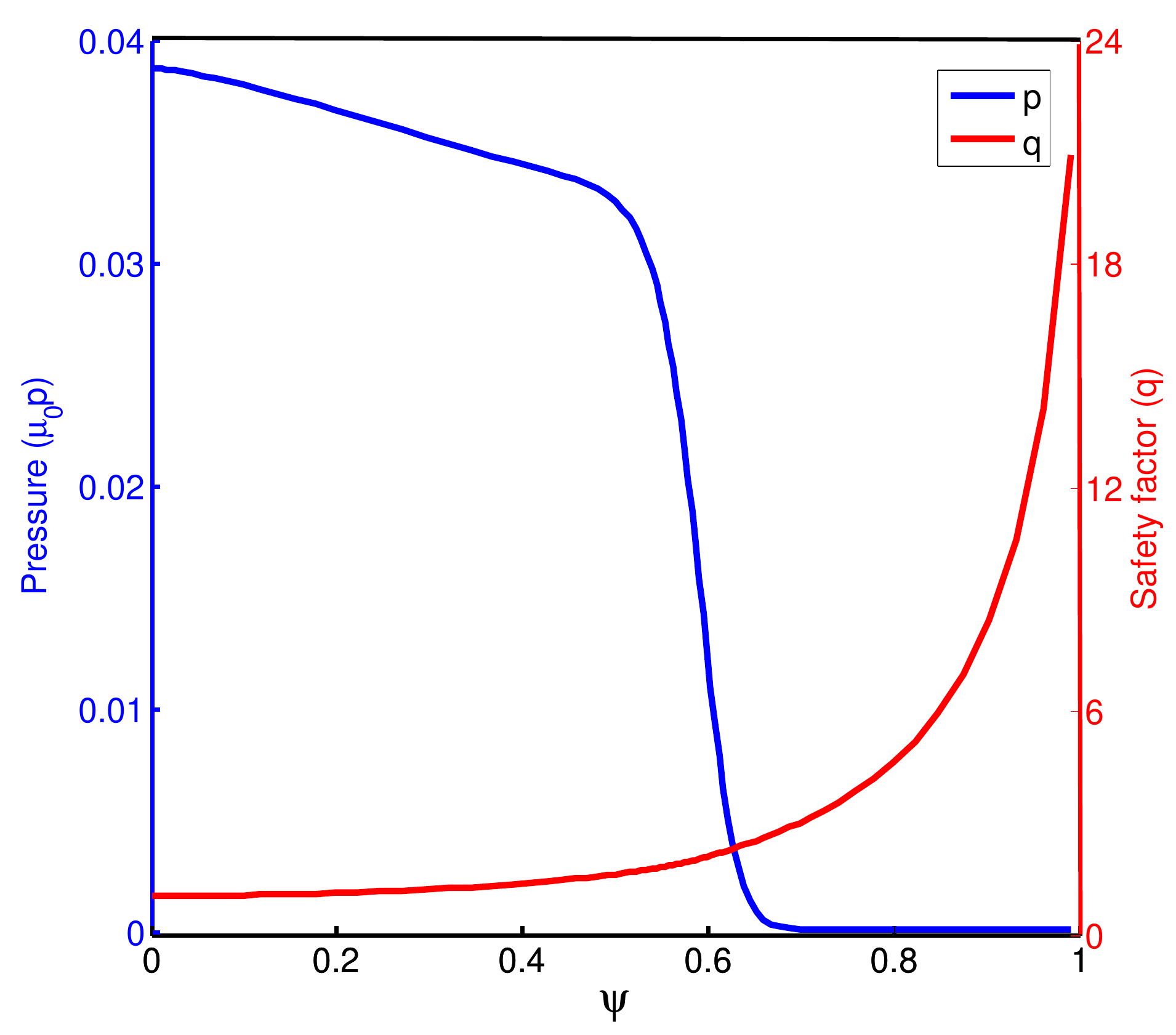}
  \put(-230,180){\textbf{(a)}}
\end{minipage}
  \begin{minipage}{0.49\textwidth}
    \includegraphics[width=1.0\textwidth,height=0.3\textheight]{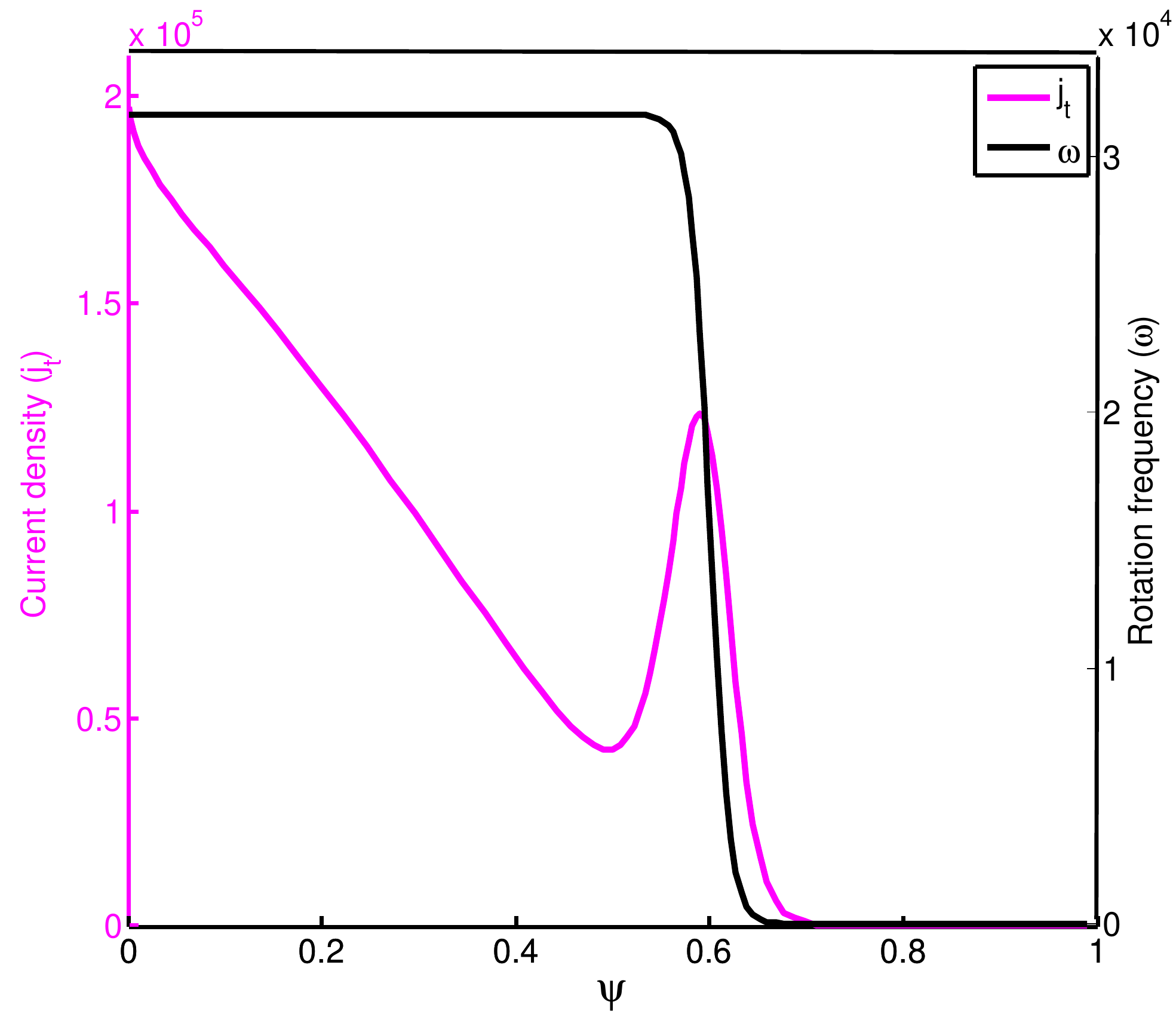}
    \put(-230,180){\textbf{(b)}}
  \end{minipage}

  \caption{Equilibrium profile as function of normalized magnetic flux $\psi$ for (a) pressure (blue line) and safety factor (red line); (b) current density (purple line) and toroidal rotation frequency (dark line).}
\label{fig:equilibrium}
\end{figure}

We consider a circular shaped tokamak equilibrium with edge pressure pedestal that has been examined in many previous studies~\citep{Miller97,Burke10,Xi12,Xia13}. The pedestal center is located at $\psi_0=0.6$ (which is marked with a blue line in Fig.~\ref{fig:equilibrium}a). It is also the peak location of the edge current profile (marked with a purple line in Fig.~\ref{fig:equilibrium}b). A toroidal flow with nonuniform radial profile is introduced to the equilibrium, where the maxmum Mach number $M=0.4$. The profile of the toroidal rotation frequency $\Omega$ is specified as 
$
\Omega(\psi)=A\{1-\tanh \left[S \left(\psi-\psi_0\right) \right]\},
$
where $A$ represents flow amplitude and $S$ determines flow shear (marked with the black line in Fig.~\ref{fig:equilibrium}b).
The maximum flow shear is set up to collocate with the center of pressure pedestal in order to allow effective interaction between the flow and edge localized modes. 

\begin{figure}[ht]
\begin{minipage}{0.49\textwidth}
  \includegraphics[width=1.0\textwidth,height=0.3\textheight]{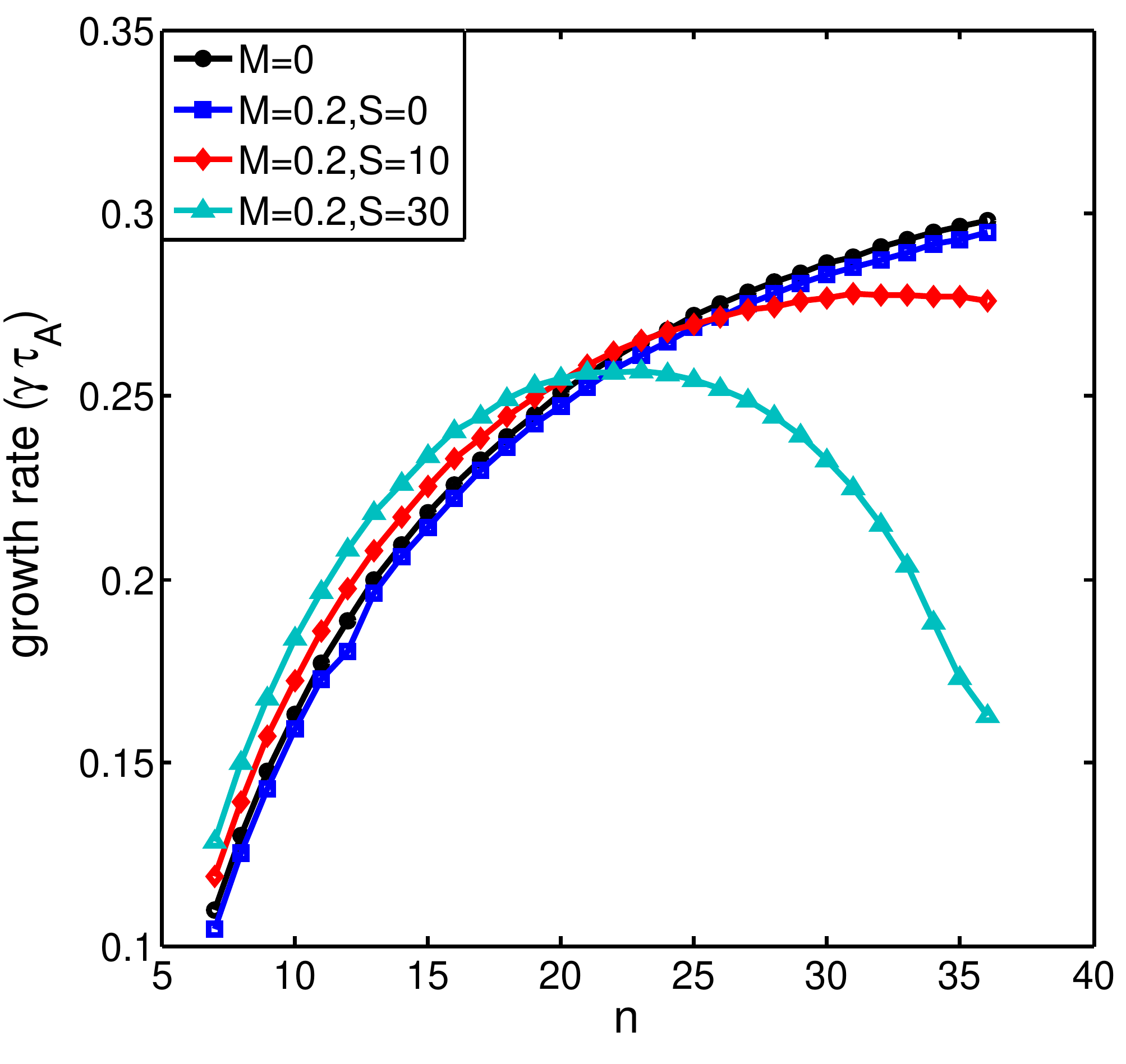}
  \put(-230,180){\textbf{(a)}}
\end{minipage}
 \begin{minipage}{0.49\textwidth}
   \includegraphics[width=1.0\textwidth,height=0.3\textheight]{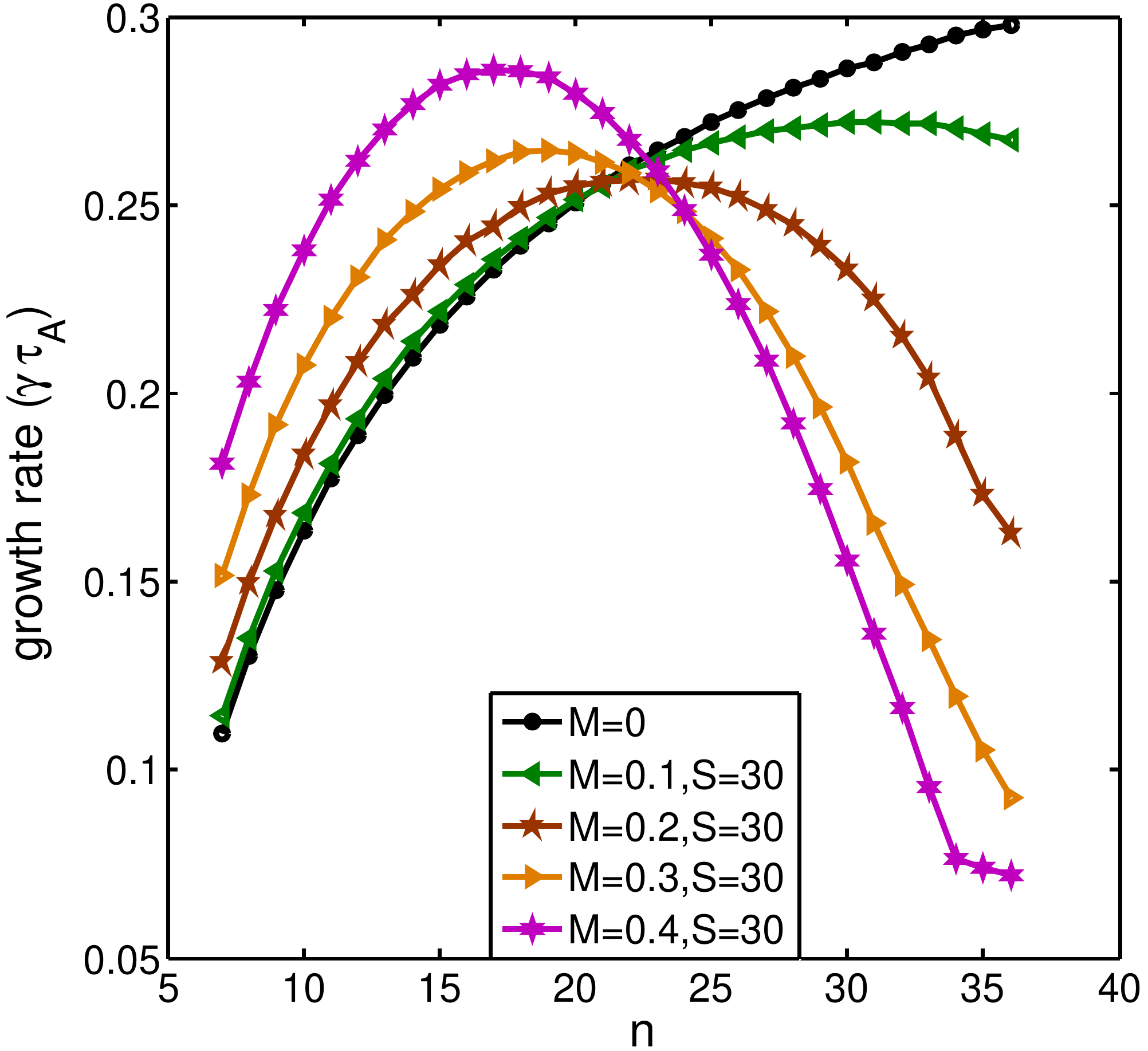}
   \put(-230,180){\textbf{(b)}}
\end{minipage}
 \caption{Linear growth rate of edge localized mode as function of toroidal mode number $n$ based on single-fluid MHD model for (a) static equilibrium ($M=0$) and equilibriums with fixed flow amplitude $M=0.2$ and different flow shear $S$; (b) static equilibrium ($M=0$) and equilibriums with fixed flow shear $S=30$ and different flow amplitude.}
\label{fig:mhd_shear_amp_growth}
\end{figure}

We first evaluate the effects of toroidal flow on the linear growth of edge localized modes for the equilibrium with fixed uniform number density, based on the single-fluid MHD model. In absence of equilibrium flow, the linear spectrum of the edge localized modes is typical of the peeling-ballooning instability in single-fluid MHD model where the growth rate monotonically increases with toroidal mode number $n$ (case $M=0$ in Fig.~\ref{fig:mhd_shear_amp_growth}a). Rigid toroidal rotation makes nearly no difference to the growth rates (case $M=0.2,S=0$ in Fig.~\ref{fig:mhd_shear_amp_growth}a). 
However, when nonuniform rotation in introduced, the flow shear effects on the growth rates become apparent, as shown in case $M=0.2,S=10$ and case $M=0.2,S=30$ in Fig.~\ref{fig:mhd_shear_amp_growth}a. The sheared flow tends to be destabilizing to the low-$n$ modes and stabilizing to the high-$n$ modes. Both effects become stronger with higher flow shear $S$, while keeping the flow amplitude parameter $A$ or the maximum Mach number $M$ fixed. Next, to evaluate the effects of flow amplitude alone, we keep the flow shear fixed ($S=30$) and vary the flow amplitude parameter $A$ or the maximum Mach number $M$. The resultant effects on growth rates are similar to those from varying flow shear alone. Higher flow amplitude gives stronger stabilization on high-$n$ modes and stronger destabilization on low-$n$ modes. However, the stabilization from sheared toroidal flow only becomes significant when the flow magnitude becomes sufficiently large (when $M>0.3$ in Fig.~\ref{fig:mhd_shear_amp_growth}b), beyond the usual range of toroidal rotation maintained in experiments. When the flow magnitude is low or moderate ($M \leq 0.3$) as in realistic tokamak experiments, the stabilizing effect appears rather weak.

\begin{figure}[ht]
\begin{minipage}{0.49\textwidth}
  \includegraphics[width=1.0\textwidth,height=0.3\textheight]{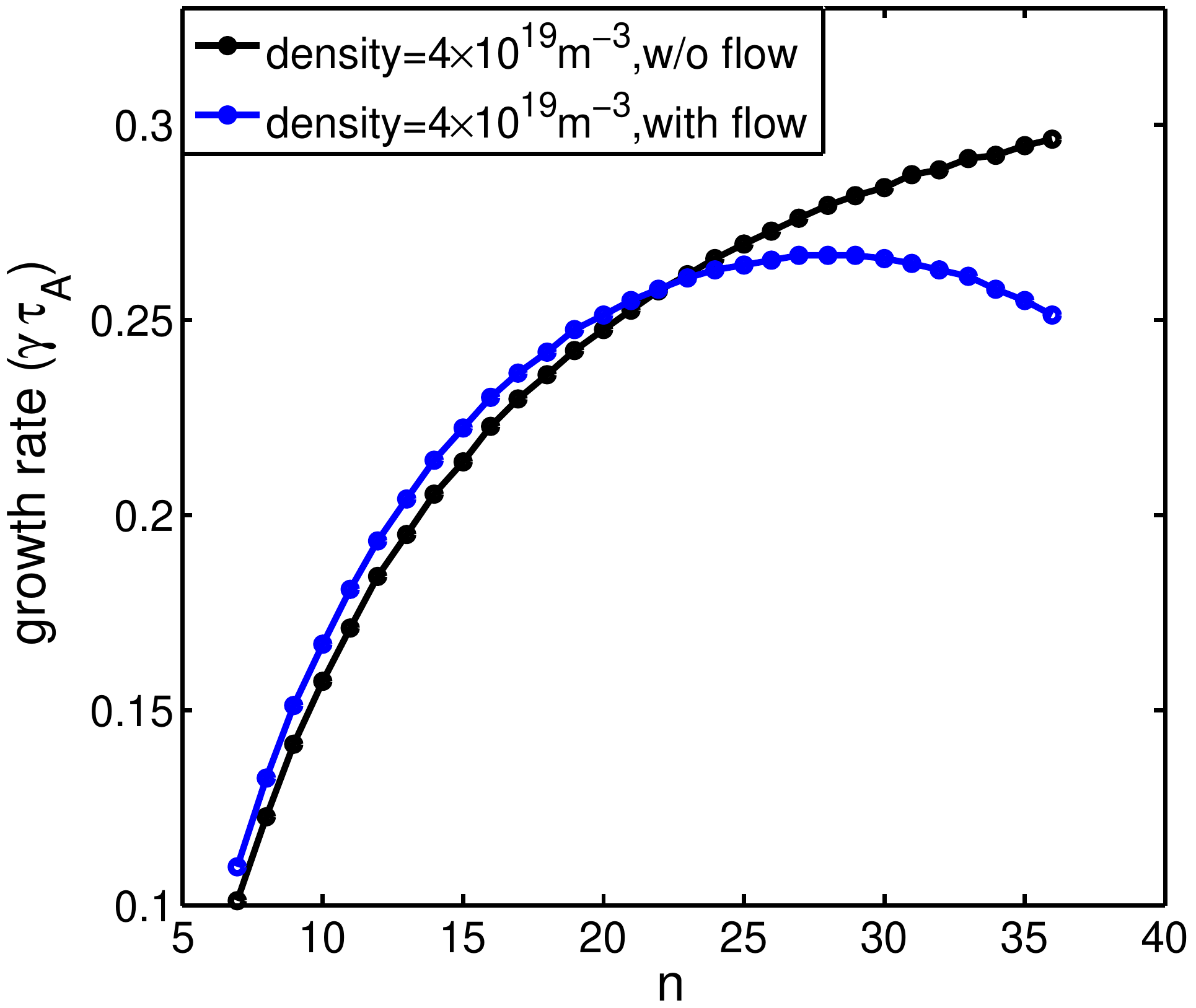}
  \put(-230,180){\textbf{(a)}}
\end{minipage}
\begin{minipage}{0.49\textwidth}
  \includegraphics[width=1.0\textwidth,height=0.3\textheight]{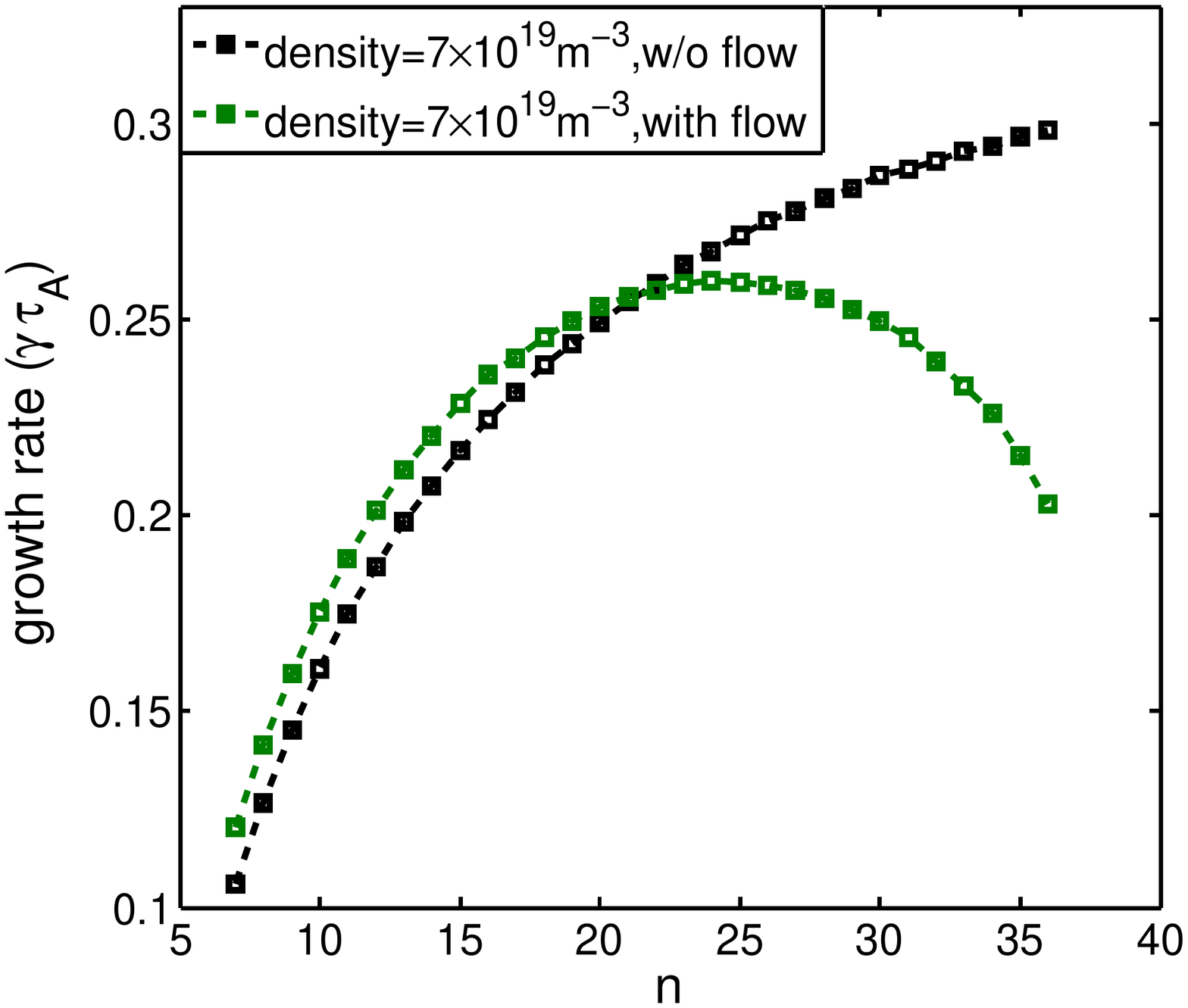}
  \put(-230,180){\textbf{(b)}}
  \end{minipage}

  \begin{minipage}{0.49\textwidth}
    \includegraphics[width=1.0\textwidth,height=0.3\textheight]{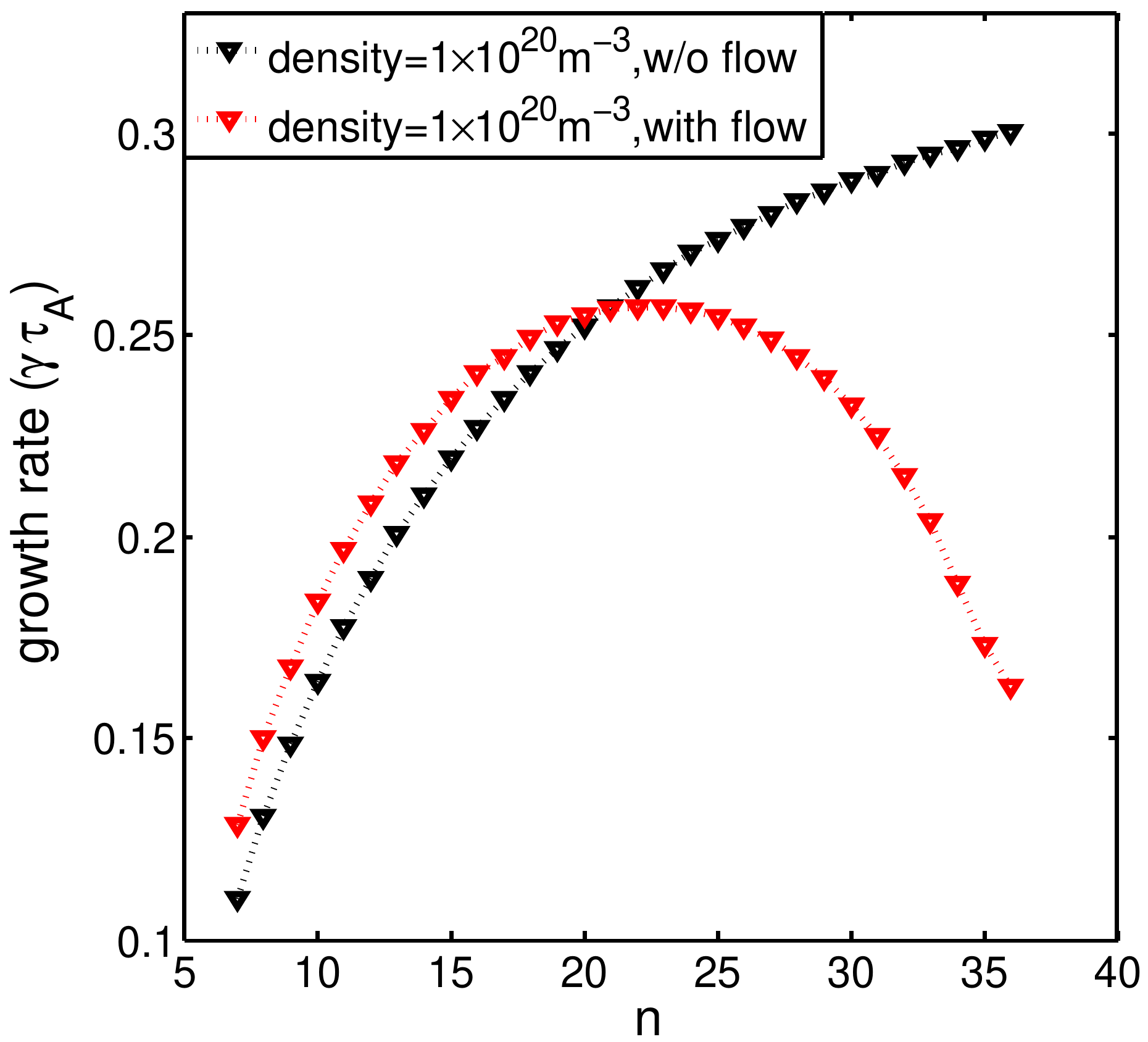}
    \put(-230,180){\textbf{(c)}}
  \end{minipage}
   \begin{minipage}{0.49\textwidth}
    \includegraphics[width=1.0\textwidth,height=0.3\textheight]{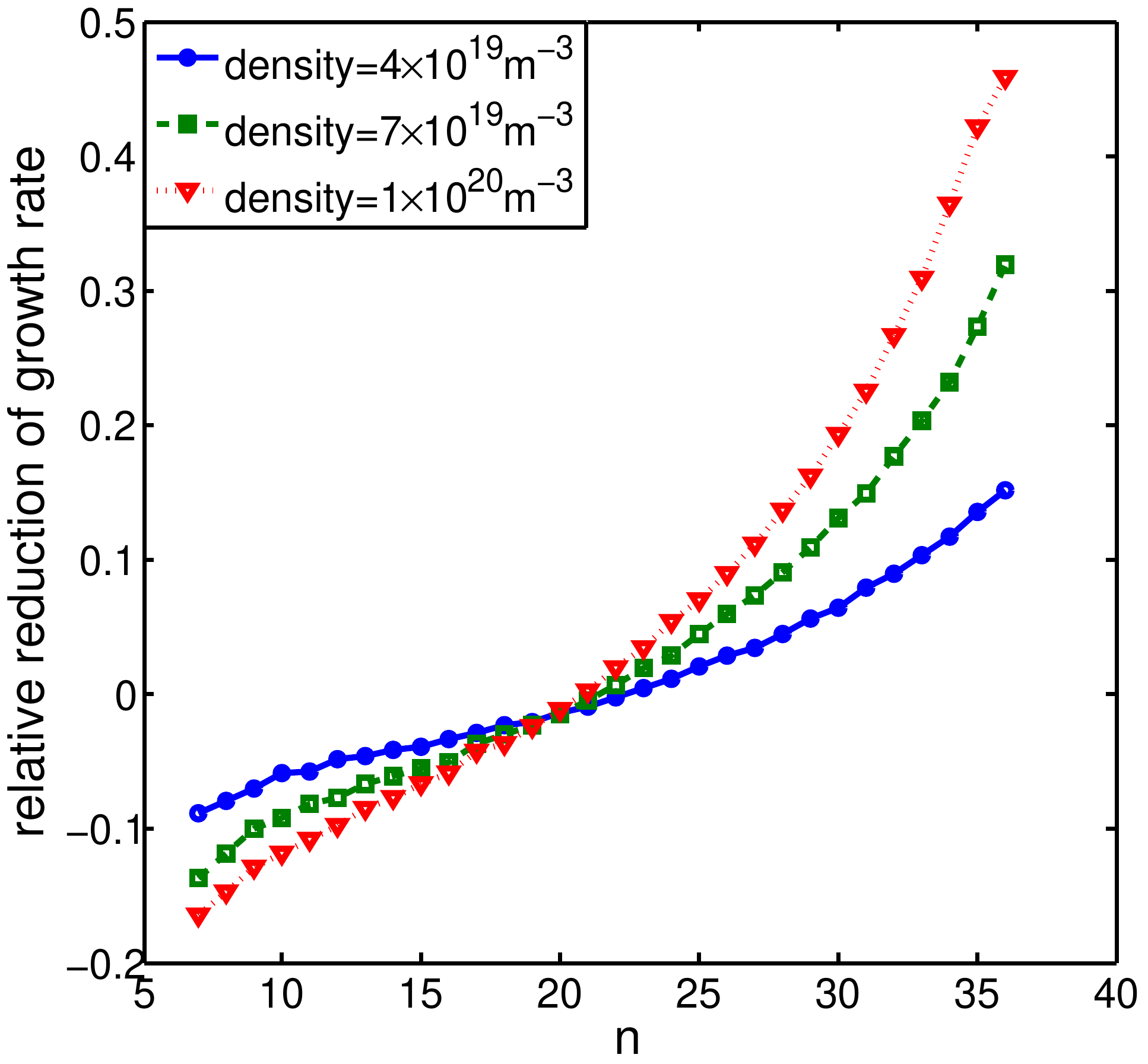}
    \put(-230,180){\textbf{(d)}}
  \end{minipage}
  \caption{Linear growth rate of edge localized mode as function of toroidal mode number $n$ based on single-fluid MHD model for equilibriums in absence of flow and in presence of a fixed flow profile ($M=0.2, S=30$) at different levels of uniform density, respectively: (a) density$=4.0\times 10^{19}m^{-3}$; (b) density$=7.0\times 10^{19}m^{-3}$; (c) density$=1.0\times 10^{20}m^{-3}$. The relative reduction in growth rate as function of toroidal mode number $n$ for each case in (a)-(c) is shown in (d).}
\label{fig:mhd_den_growth}
\end{figure}

The first new finding from our calculations is that the toroidal flow stabilizing effect can be enhanced by increasing plasma number density, even when flow magnitude is moderate or low ($M=0.2$). To demonstrate this effect of plasma density, we keep the toroidal flow profile fixed with $M=0.2$ and $S=30$. Increasing the uniform density value from $4.0\times 10^{19} m^{-3}$ (Fig.~\ref{fig:mhd_den_growth}a) to $7.0\times 10^{19} m^{-3}$ (Fig.~\ref{fig:mhd_den_growth}b) and then $1.0\times 10^{20} m^{-3}$ (Fig.~\ref{fig:mhd_den_growth}c) monotonically decreases the growth rates of high-$n$ modes while resulting in nearly no change of low-$n$ modes (Fig.~\ref{fig:mhd_den_growth}d). The normalized growth rate of $n=36$ mode is reduced by more than $46\%$ in the presence of sheared flow with the higher plasma density of $1.0\times 10^{20} m^{-3}$, which is substantially more than the relative reduction in growth rate ($18\%$) in the presence of sheared flow with the lower plasma density of $4.0\times 10^{19} m^{-3}$. Thus high-$n$ modes can be strongly stabilized by low to moderate flow in combination with  increased number density hence collisionality.

\begin{figure}[ht]
\begin{minipage}{0.49\textwidth}
  \includegraphics[width=1.0\textwidth,height=0.3\textheight]{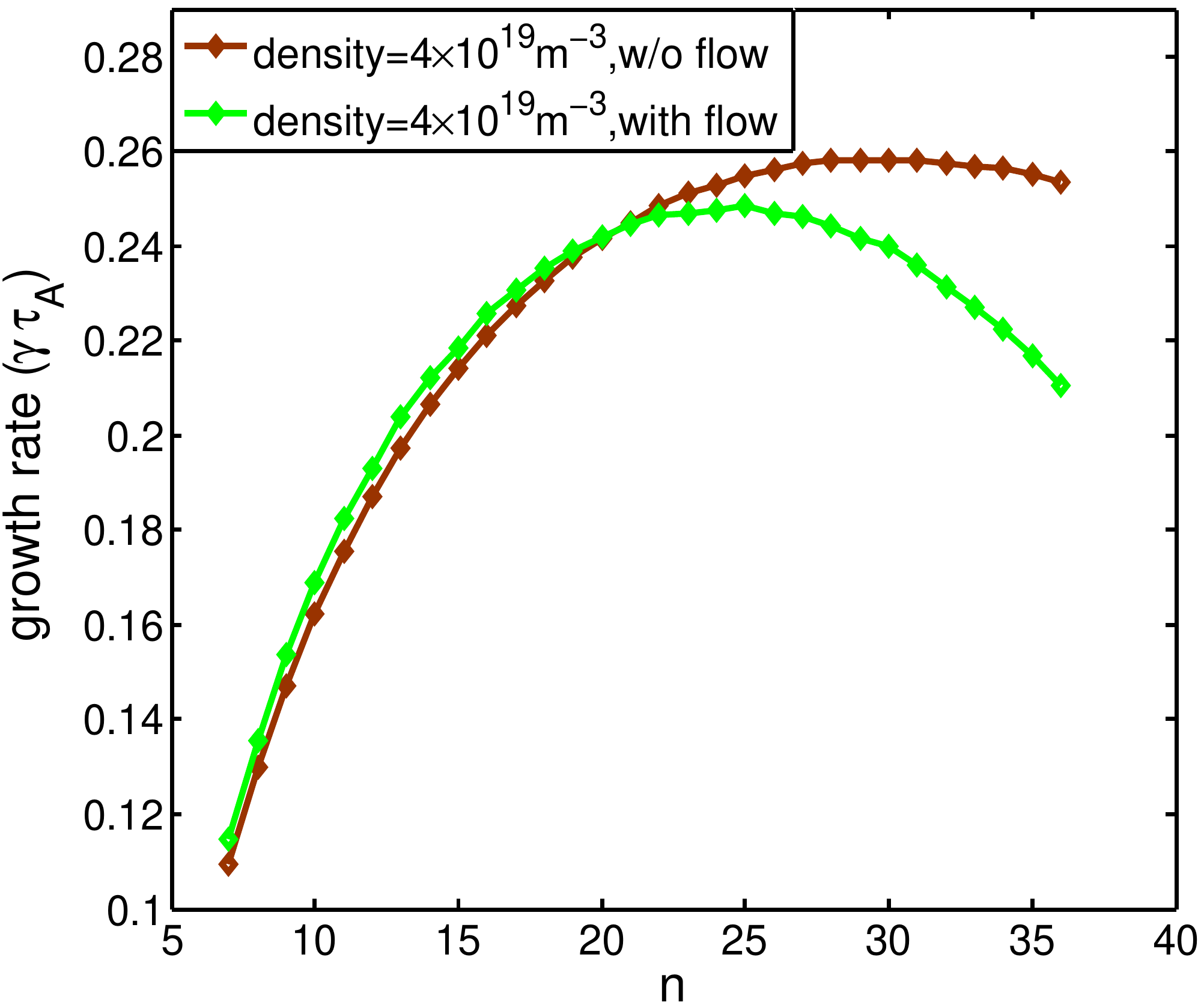}
  \put(-230,180){\textbf{(a)}}
\end{minipage}
\begin{minipage}{0.49\textwidth}
  \includegraphics[width=1.0\textwidth,height=0.3\textheight]{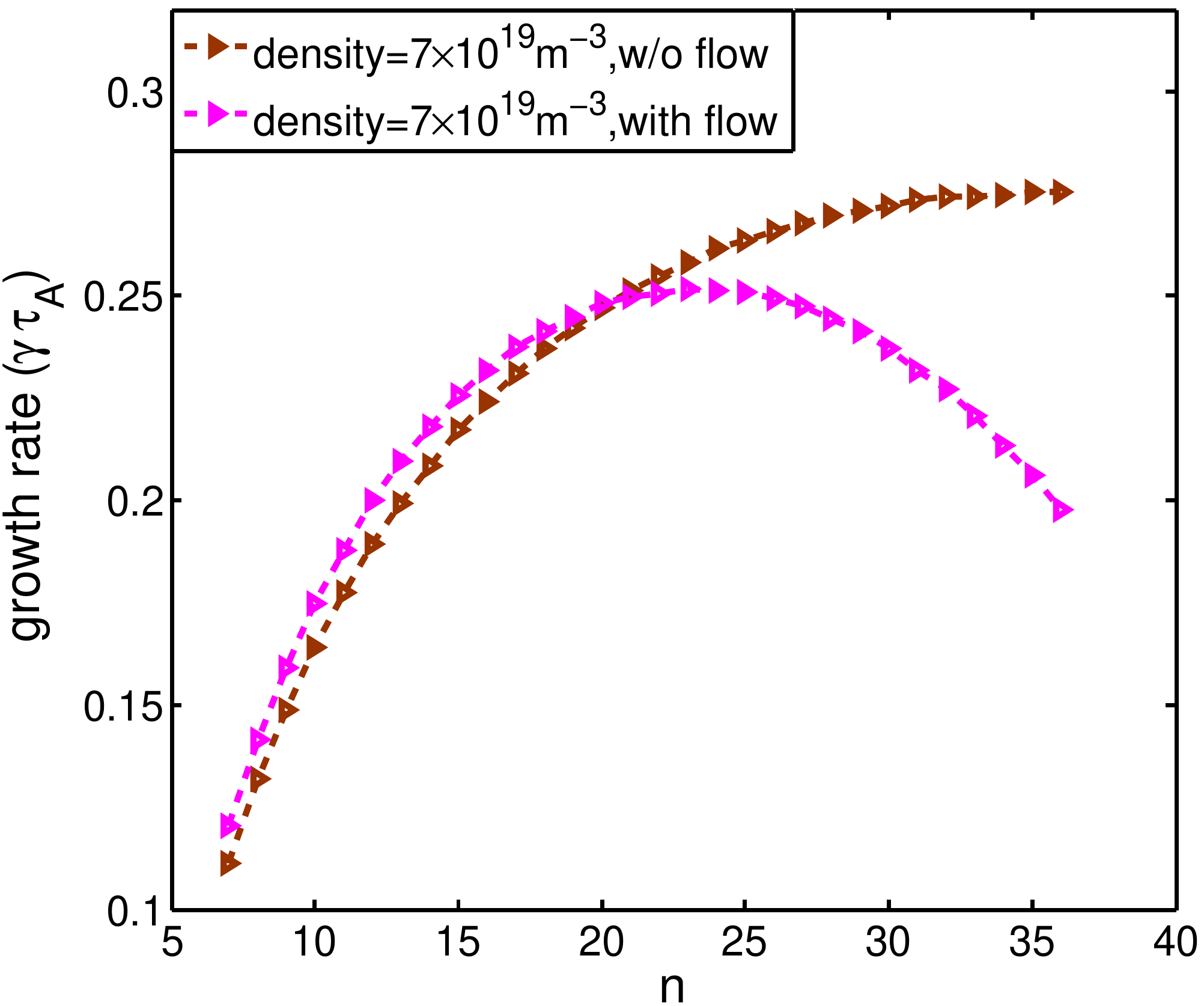}
  \put(-230,180){\textbf{(b)}}
  \end{minipage}

  \begin{minipage}{0.49\textwidth}
    \includegraphics[width=1.0\textwidth,height=0.3\textheight]{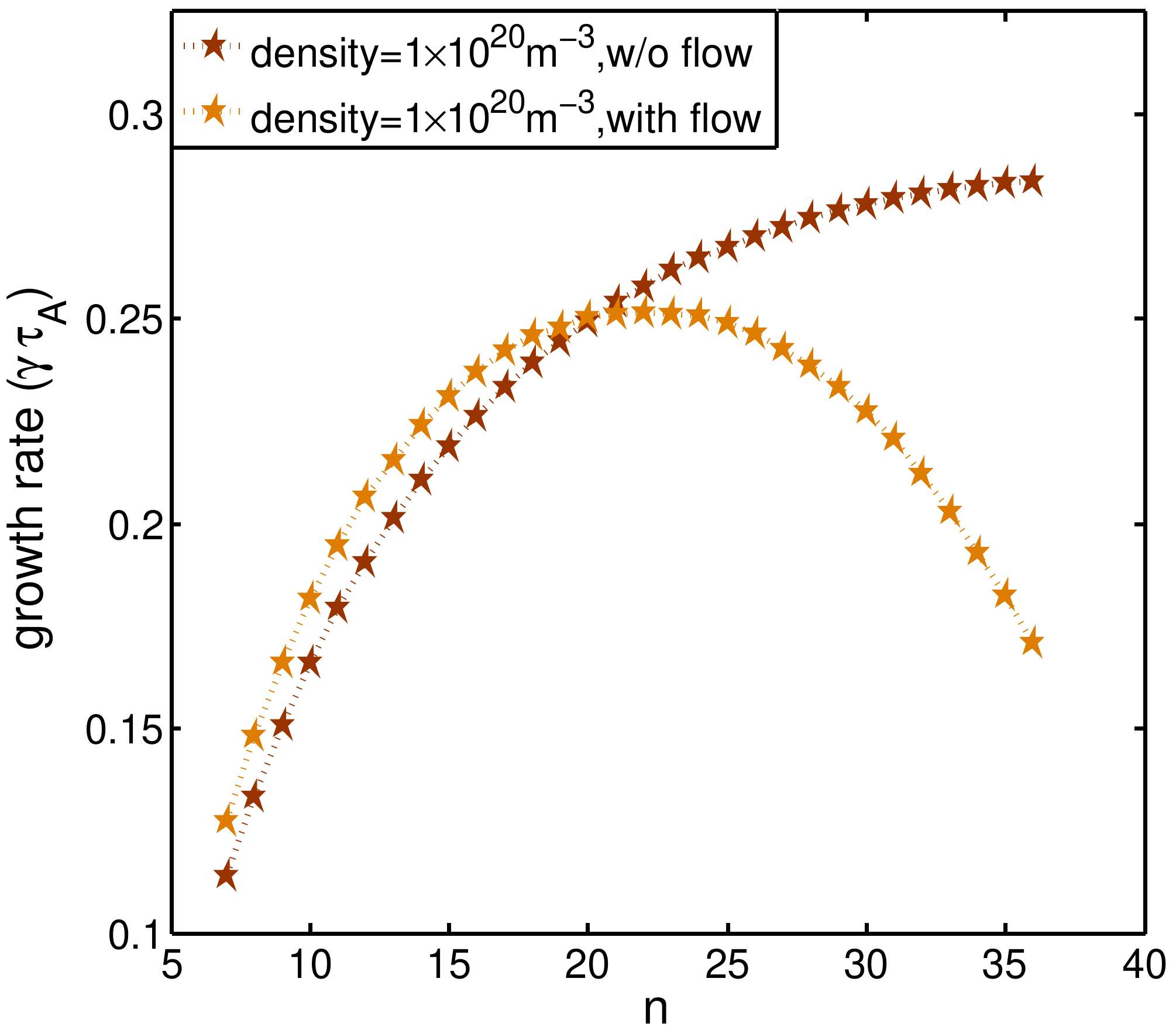}
    \put(-230,180){\textbf{(c)}}
  \end{minipage}
    \begin{minipage}{0.49\textwidth}
    \includegraphics[width=1.0\textwidth,height=0.3\textheight]{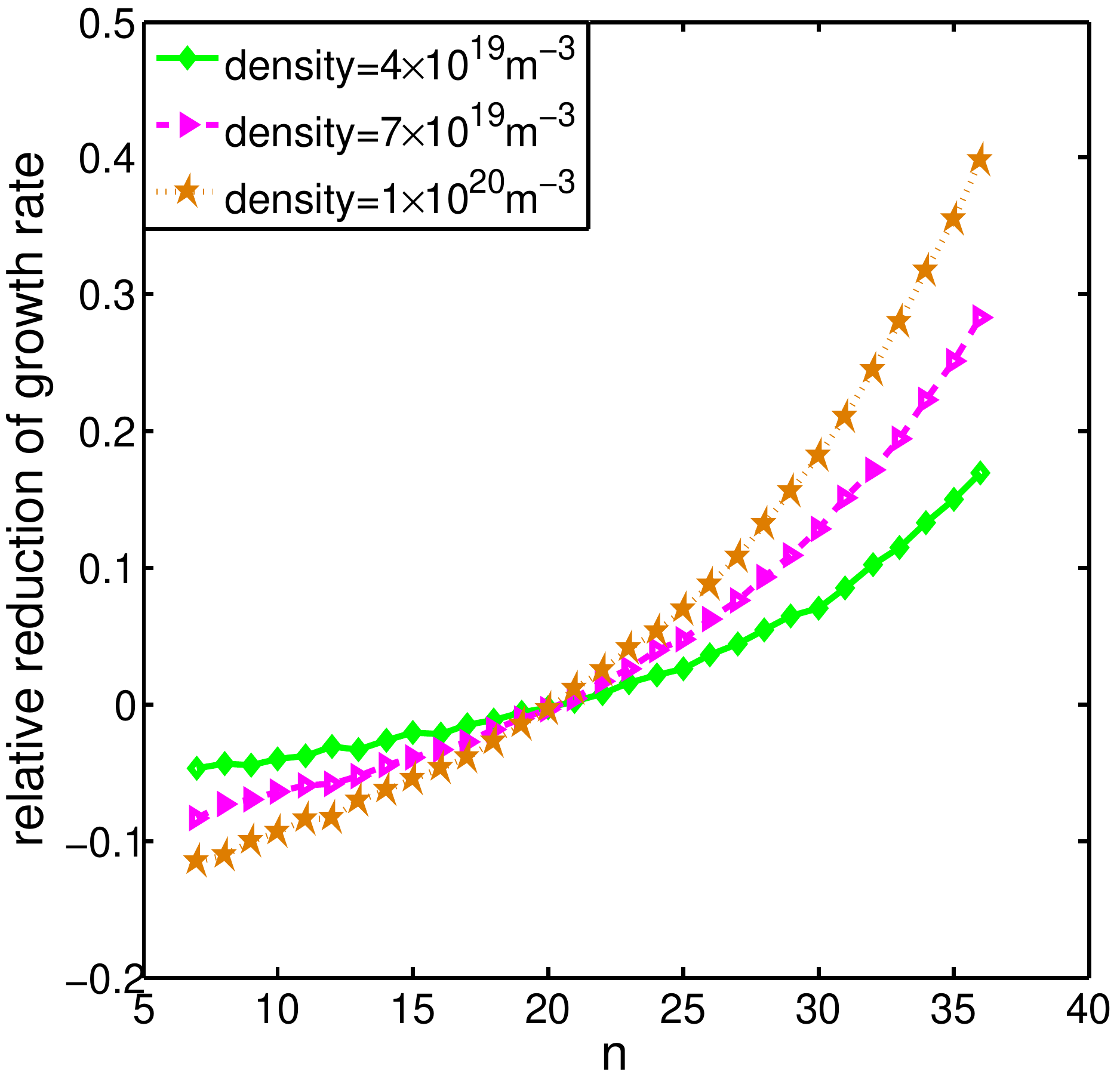}
    \put(-230,180){\textbf{(d)}}
  \end{minipage}
\caption{Linear growth rate of edge localized mode as function of toroidal mode number $n$ based on two-fluid MHD model for equilibriums in absence of flow and in presence of a fixed flow profile ($M=0.2, S=30$) at different levels of uniform density, respectively: (a) density$=4.0\times 10^{19}m^{-3}$; (b) density$=7.0\times 10^{19}m^{-3}$; (c) density$=1.0\times 10^{20}m^{-3}$. The relative reduction in growth rate as function of toroidal mode number $n$ for each case in (a)-(c) is shown in (d).}
\label{fig:2fl_den_growth}
\end{figure}

To check the generality of our findings, we turn to the full extended MHD model including two-fluid and finite-Larmor-radius (FLR) effects. For the static equilibriums, the increased density appears to be destabilize to high-$n$ modes, which was also previously reported in~\citep{King16}.
Then we consider the rotating equilibrium cases with uniform density. It turns out that in the presence of sheared flow, increasing plasma density can be slightly destabilizing to the low $n$ modes while strongly stabilizing to the high-$n$ modes (Fig.~\ref{fig:2fl_den_growth}a, b and c). This is similar to the flow and density effects on ELMs from the single-fluid MHD calculations reported in previous paragraph. In another words, the same combined effects of density and toroidal rotation on ELM growth rates persist even in the two-fluid MHD model. Similarly, this combined effect may be better demonstrated in Fig.~\ref{fig:2fl_den_growth}d, where the relative reduction in growth rate of $n=36$ mode is $19\%$ at plasma density of $4.0\times 10^{19} m^{-3}$, and becomes much more pronounced ($40\%$) at plasma density of $1.0\times 10^{20} m^{-3}$.

\begin{figure}[ht]
\begin{minipage}{0.49\textwidth}
  \includegraphics[width=1.0\textwidth,height=0.32\textheight]{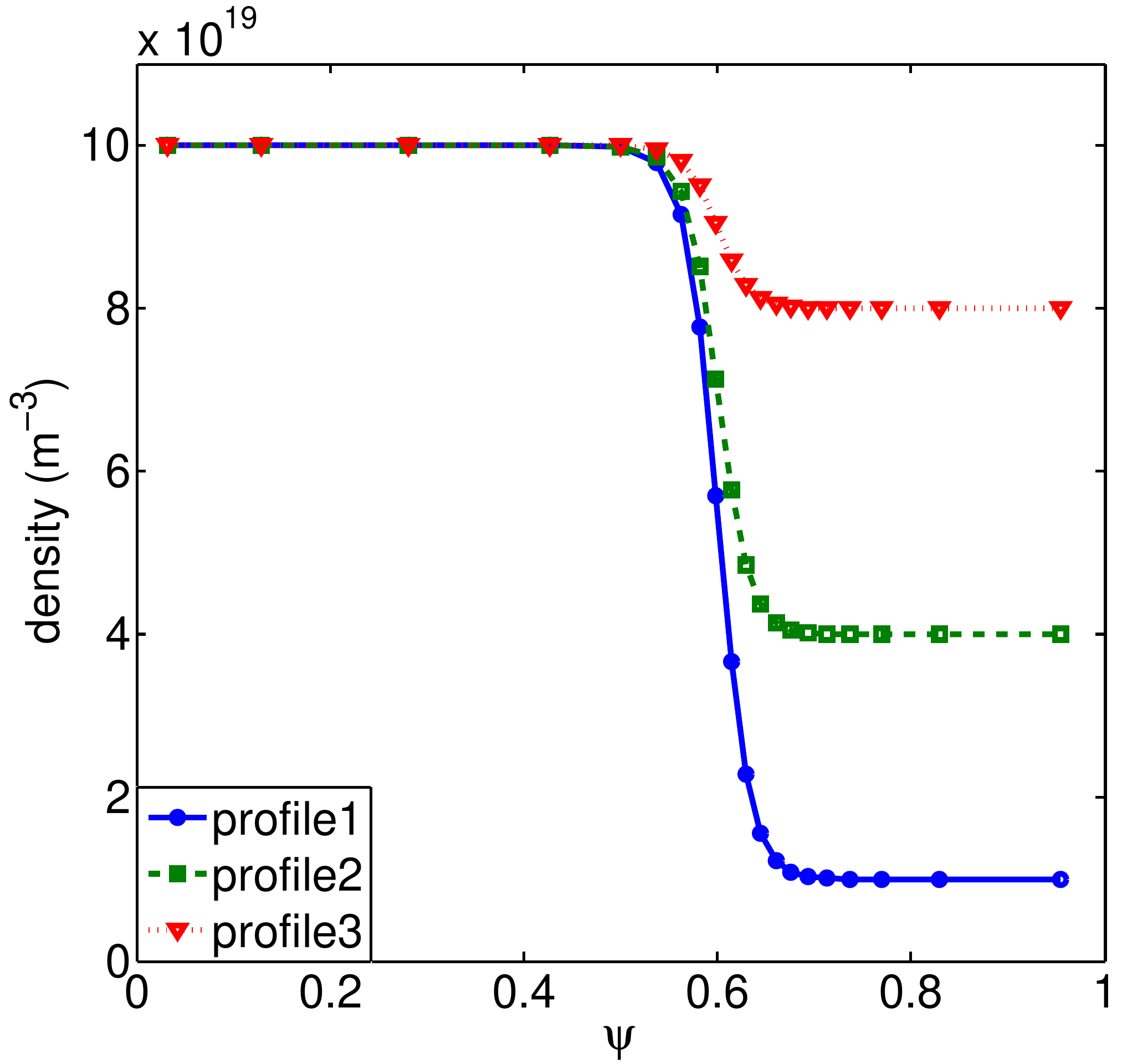}
  \put(-230,220){\textbf{(a)}}
\end{minipage}
\begin{minipage}{0.49\textwidth}
  \includegraphics[width=1.0\textwidth,height=0.32\textheight]{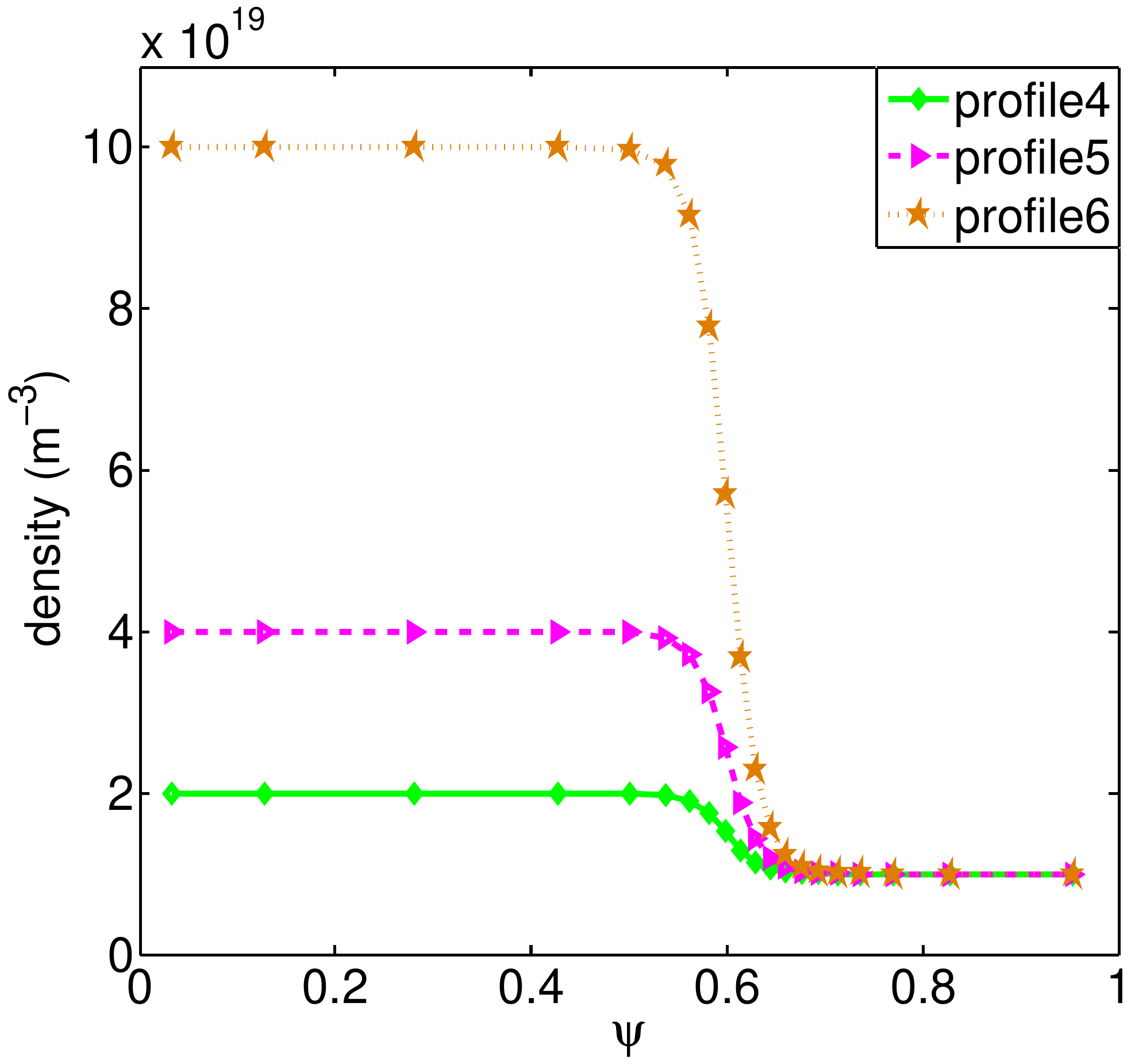}
  \put(-230,220){\textbf{(b)}}
\end{minipage}
  \caption{Equilibriums with non-uniform density profiles where (a) core density level is fixed ($1.0\times 10^{20} m^{-3})$ and edge density levels are different; (b) edge density level is fixed ($1.0\times 10^{19} m^{-3})$ and core density levels are different.}
\label{fig:den_prof}
\end{figure}

\begin{figure}[ht]
\begin{minipage}{0.49\textwidth}
  \includegraphics[width=0.8\textwidth,height=0.2\textheight]{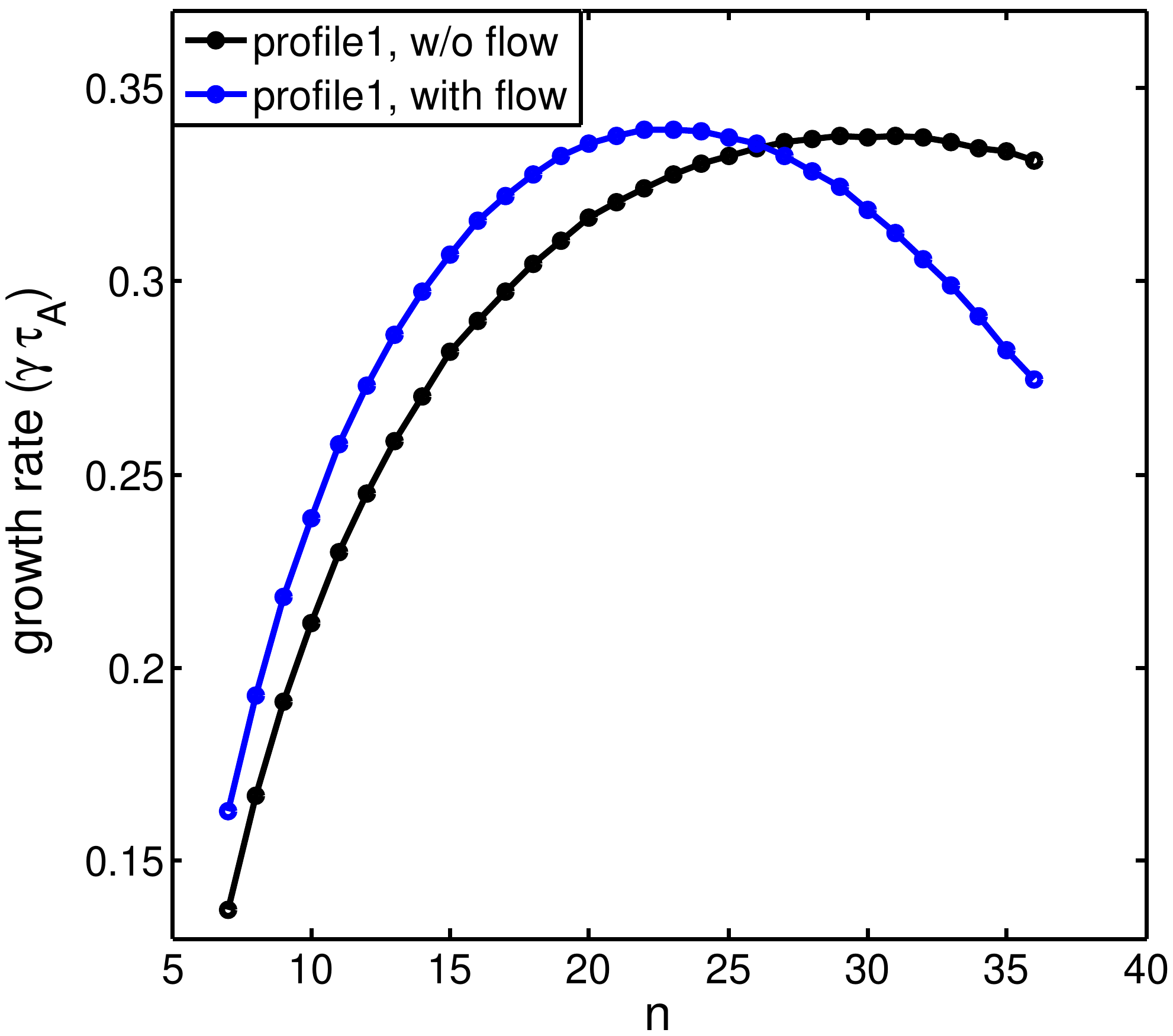}
  \put(-200,120){\textbf{(a)}}
\end{minipage}
\begin{minipage}{0.49\textwidth}
  \includegraphics[width=0.8\textwidth,height=0.2\textheight]{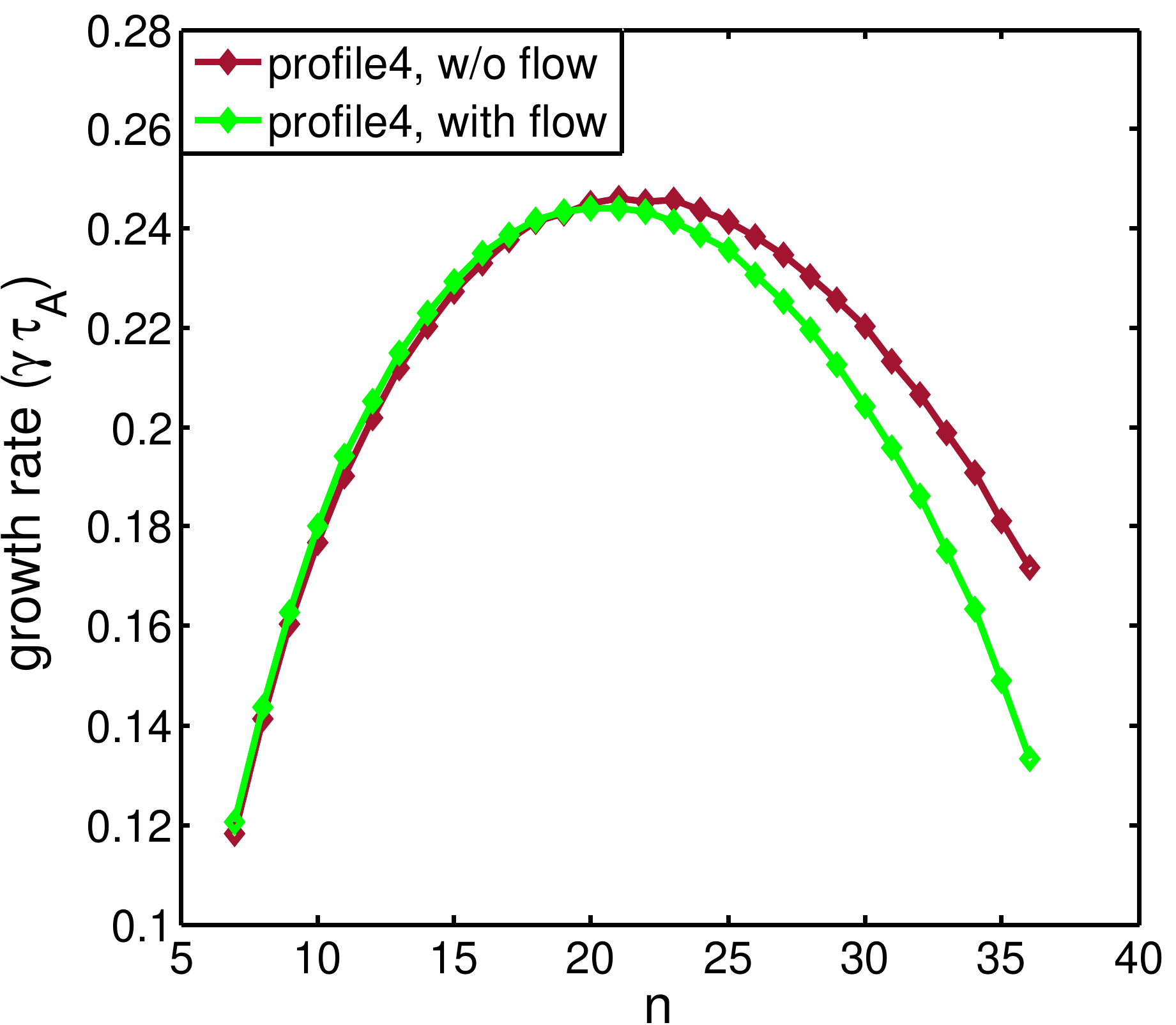}
  \put(-200,120){\textbf{(e)}}
\end{minipage}

\begin{minipage}{0.49\textwidth}
  \includegraphics[width=0.8\textwidth,height=0.2\textheight]{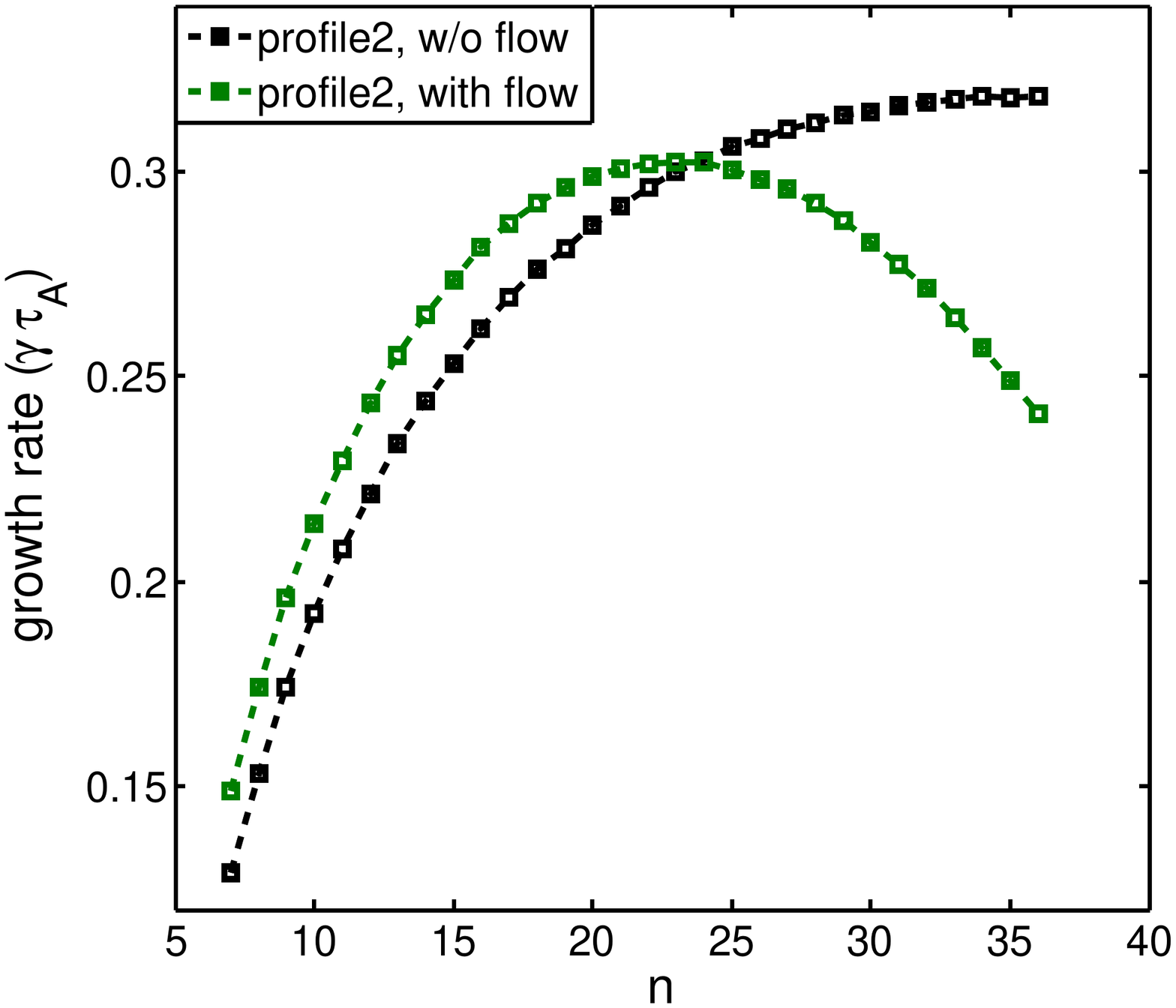}
  \put(-200,120){\textbf{(b)}}
\end{minipage}
\begin{minipage}{0.49\textwidth}
  \includegraphics[width=0.8\textwidth,height=0.2\textheight]{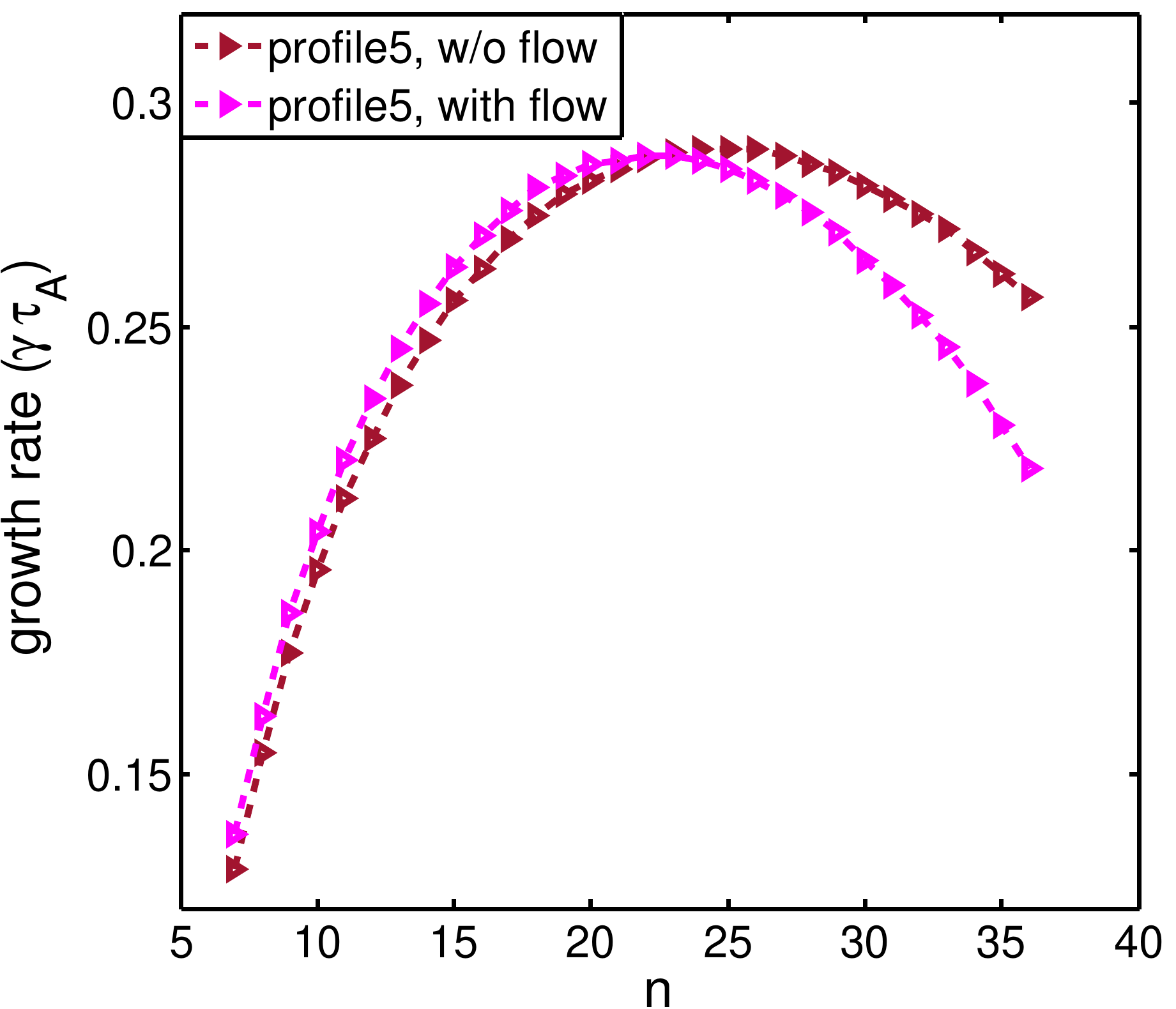}
  \put(-200,120){\textbf{(f)}}
\end{minipage}

\begin{minipage}{0.49\textwidth}
  \includegraphics[width=0.8\textwidth,height=0.2\textheight]{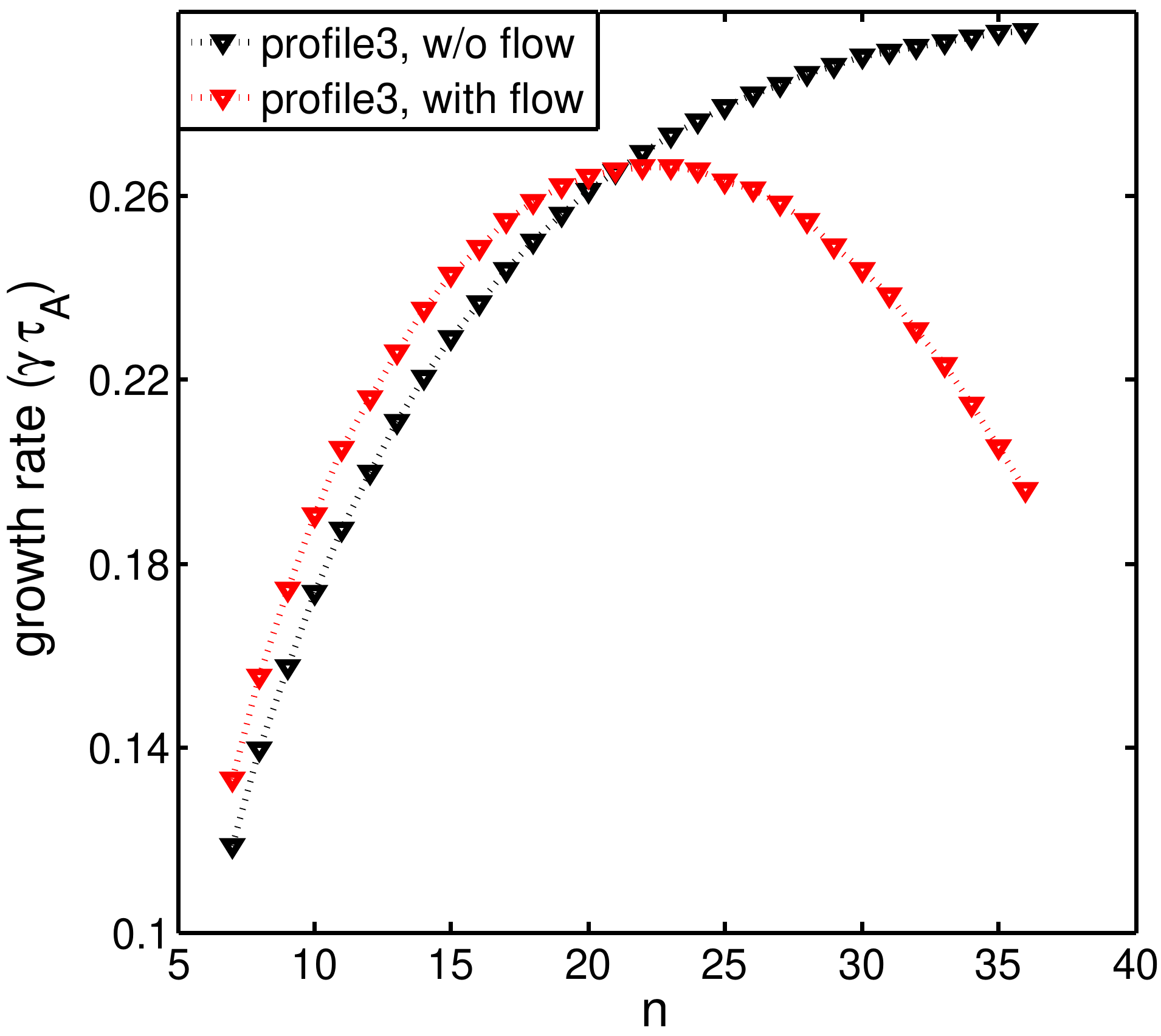}
  \put(-200,120){\textbf{(c)}}
\end{minipage}
\begin{minipage}{0.49\textwidth}
  \includegraphics[width=0.8\textwidth,height=0.2\textheight]{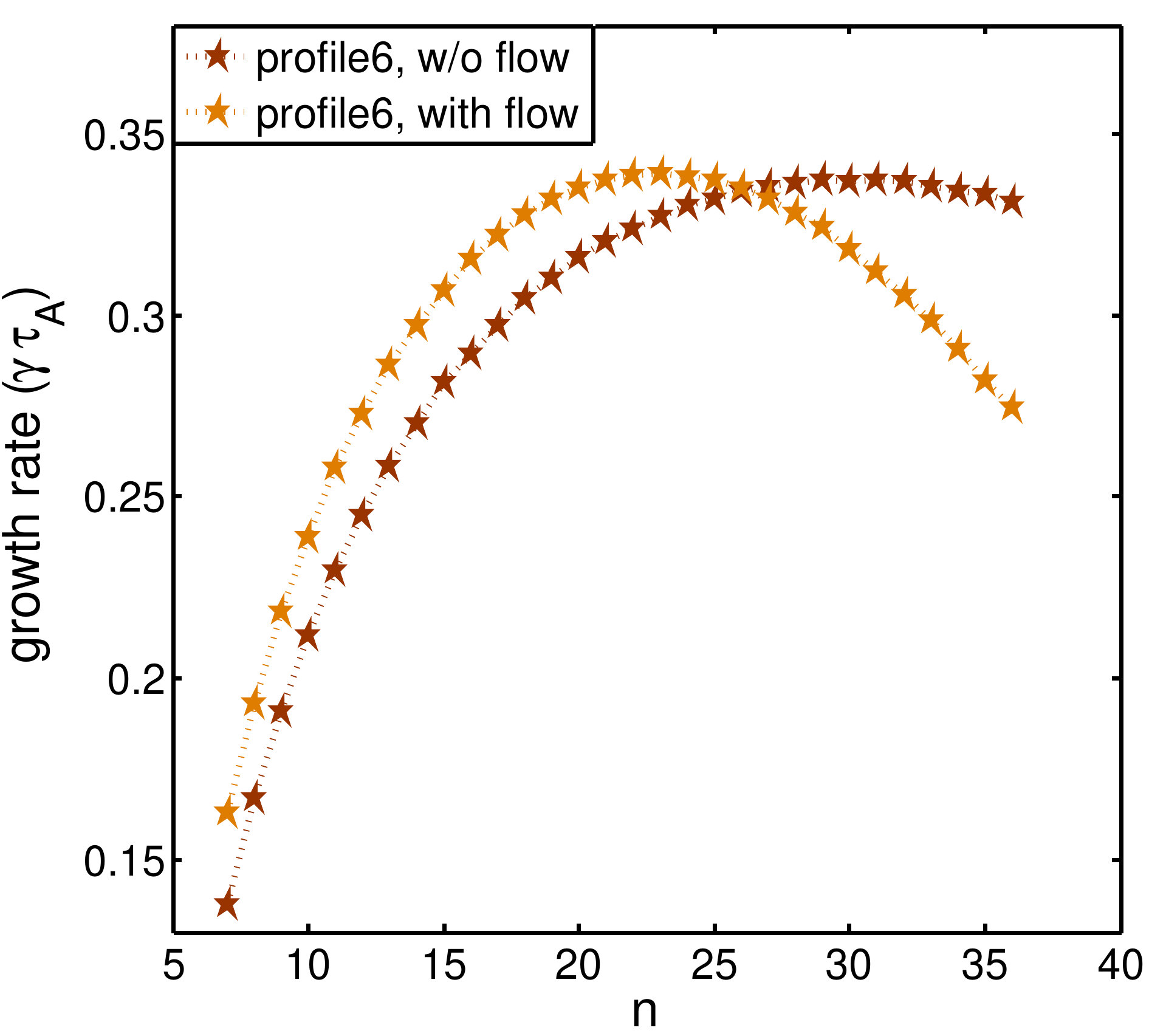}
  \put(-200,120){\textbf{(g)}}
\end{minipage}

\begin{minipage}{0.49\textwidth}
  \includegraphics[width=0.8\textwidth,height=0.2\textheight]{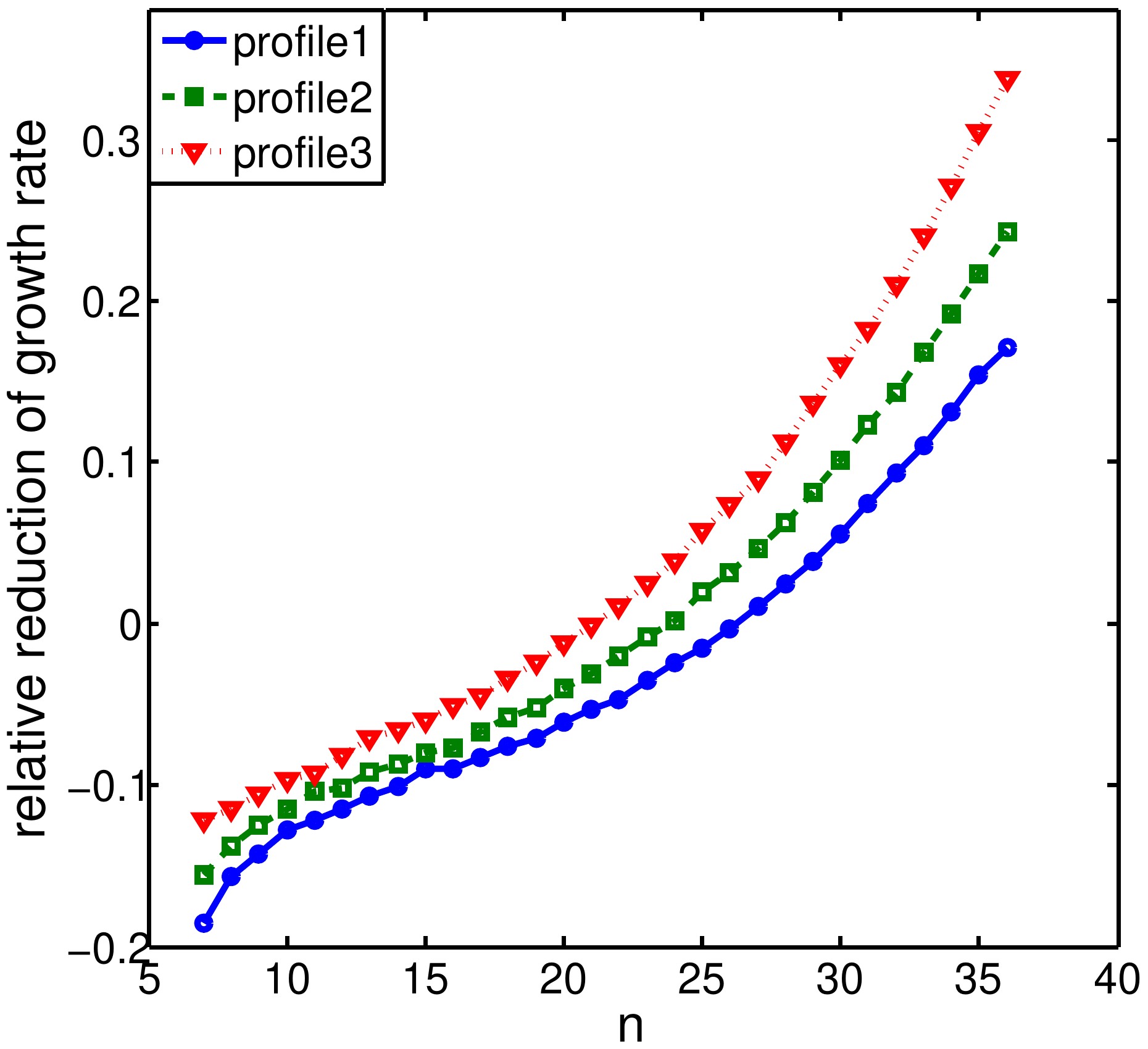}
  \put(-200,120){\textbf{(d)}}
\end{minipage}
\begin{minipage}{0.49\textwidth}
  \includegraphics[width=0.8\textwidth,height=0.2\textheight]{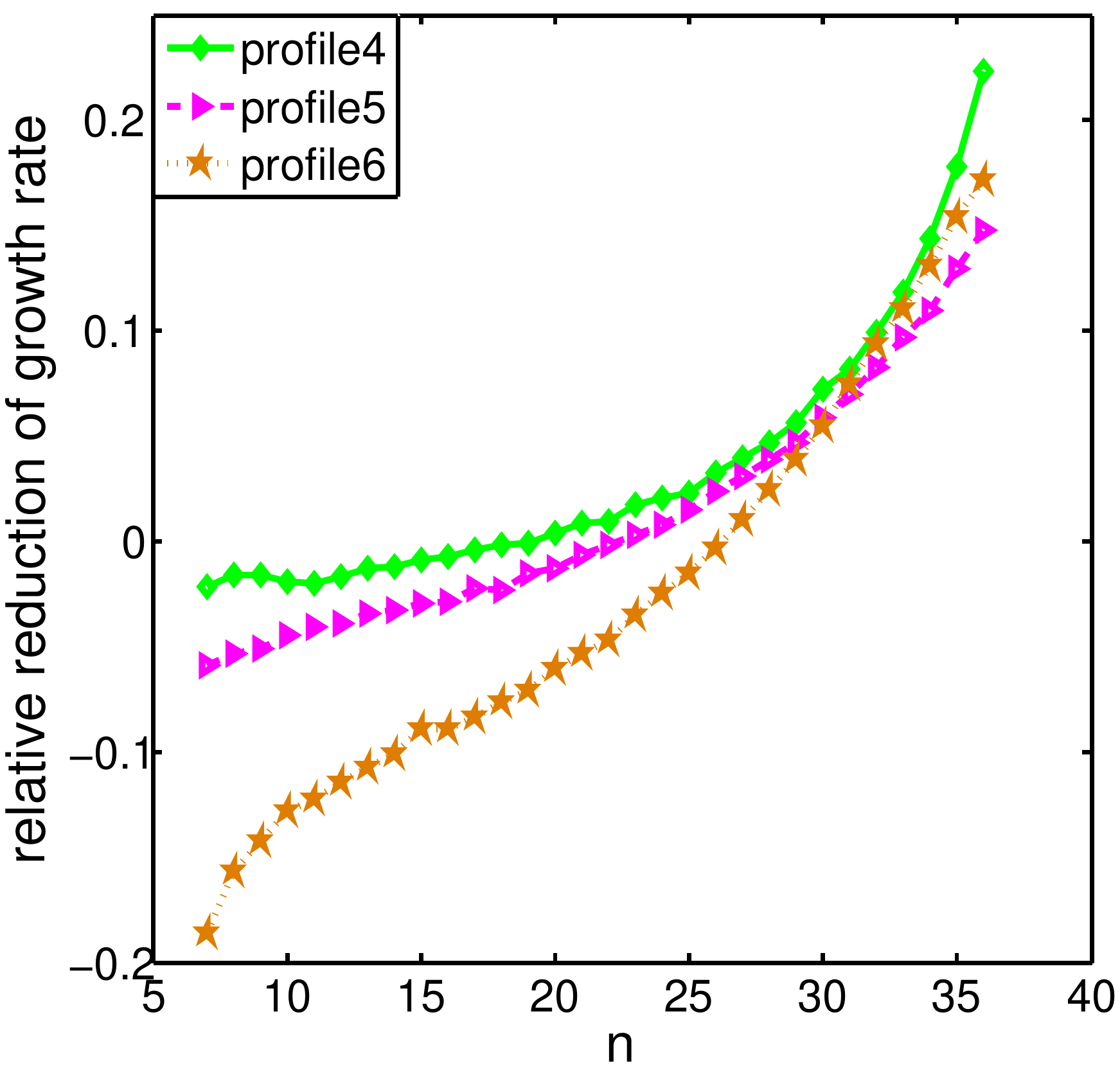}
  \put(-200,120){\textbf{(h)}}
\end{minipage}

\caption{Linear growth rate of edge localized mode as function of toroidal mode number $n$ based on two-fluid MHD model for equilibriums in absence of flow and in presence of a fixed toroidal flow profile ($M=0.2$ and $S=30$), respectively: (a)-(c) stands for the cases of density ``profile 1'' to density ``profile 3'' from Fig.~\ref{fig:den_prof}a and (e)-(g) stands for density ``profile 4'' to density ``profile 6'' from Fig.~\ref{fig:den_prof}b. The corresponding relative reductions in growth rates are shown in (d) and (h), respectively.}
\label{fig:2fl_den_growth_pro}
\end{figure}

In reality, the radial profile of plasma density in a tokamak is not uniform. So we now address the question whether or how the non-uniformly distributed density may still enhance the stabilization of ELMs by toroidal flow. The radial profile of plasma density in a H-mode tokamak can be modeled as a composition of an edge pedestal region with steep gradient along with a wide flat-top core region. For our analysis, the density profile covering both core and edge regions is modeled with hyperbolic tangent function (Fig.~\ref{fig:den_prof}). The location of density pedestal is set to be same as the pressure pedestal at $\psi$=0.6. At the first step, we keep the density of core region fixed ($\simeq 1.0\times 10^{20} m^{-3}$) and increase the edge density level from $1.0\times 10^{19} m^{-3}$ to $8.0\times 10^{19} m^{-3}$ sequentially (Fig.~\ref{fig:den_prof}a). We then keep the edge density constant ($\simeq 1.0\times 10^{19} m^{-3}$), and vary the density of core region from $2.0\times 10^{19} m^{-3}$ to $1.0\times 10^{20} m^{-3}$ (Fig.~\ref{fig:den_prof}b). As it turns out, increasing the density of edge region enhances the stabilizing effects of flow on high-$n$ modes while making nearly no change to low-$n$ modes (Figs.~\ref{fig:2fl_den_growth_pro}a, b and c). On the other hand, increasing the core region density enhances the destabilizing effects of flow on the $n\leq 30$ modes while resulting in nearly no effect on the $n > 30$ modes (Figs.~\ref{fig:2fl_den_growth_pro}e, f and g). These effects are summarized in Figs.~\ref{fig:2fl_den_growth_pro}d and h, which show the relative reduction of growth rate from profiles $1-3$ and profiles $4-6$. Comparing the two sets of results reveals that it is the edge region density instead of the core region density that can substantially enhance the stabilizing effect on ELMs from toroidal sheared flow. Such a finding is consistent with recent experimental observations, where gas puffing was used to increase edge density and collisionality. In those experiments, ELM mitigation and suppression are found to be more favorably achieved in those higher density/collisionality regimes for a given toroidal sheared flow~\citep{Oyama10,Xu.Ma.Li14,Zweben14,Kong}.

\begin{figure}[ht]
  \includegraphics[width=0.6\textwidth,height=0.35\textheight]{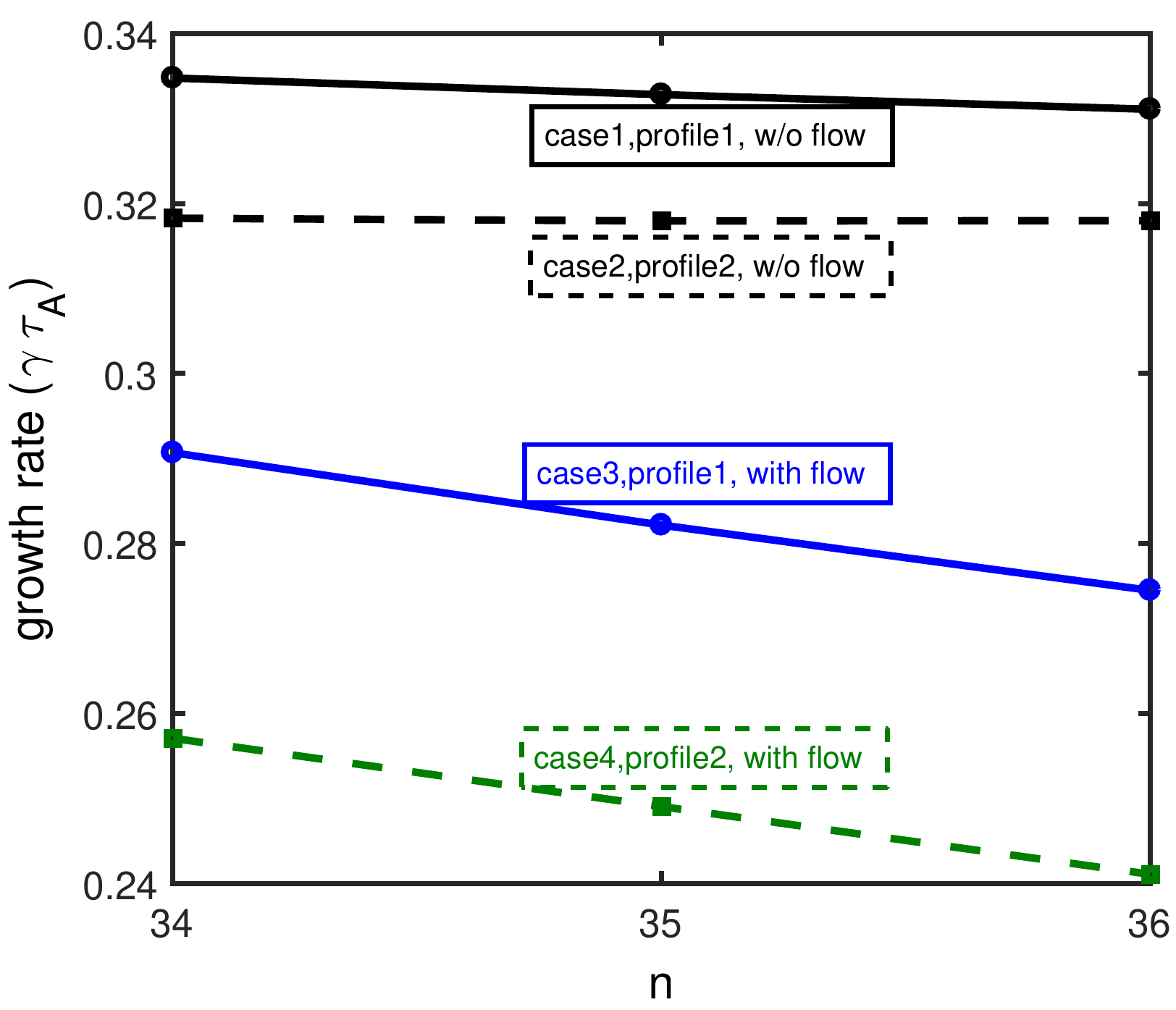}
  \caption{Linear growth rates of toroidal modes $n=34-36$ for two sets of equilibriums with the same core density level of $1.0 \times10^{20}m^{-3}$ and two different edge density values of $1.0 \times 10^{19}m^{-3}$ and $4.0 \times 10^{19}m^{-3}$ (denoted as ``profile 1'' and ``profile 2'' respectively, same as those in Fig.~\ref{fig:den_prof}a). For each set of density profile, the static equilibrium and the equilibrium in presence of a same toroidal sheared flow ($M=0.2$ and $S=30$) are considered.}
\label{fig:growth_flow_density}
\end{figure}

In passing, it is worth pointing out that the combined stabilizing effects of density and toroidal flow is not merely a simple addition of the separate effects from density and flow alone. To demonstrate this, we consider two non-uniform density profiles with the same core density level of $1.0\times 10^{20}m^{-3}$ and two different edge density values of $1.0 \times 10^{19}m^{-3}$ (denoted as ``profile 2 '' in Figs.~\ref{fig:den_prof}a and ~\ref{fig:growth_flow_density}) and $4.0 \times 10^{19}m^{-3}$ (denoted as ``profile 1'' in Figs.~\ref{fig:den_prof}a and ~\ref{fig:growth_flow_density}) respectively. We compare the growth rates of ELMs in these two cases with and without toroidal flow. For illustration purpose, we show the growth rates of modes with toroidal mode numbers from $n=34$ to $n=36$ in Fig.~\ref{fig:growth_flow_density}. First, we consider the density effects alone in absence of flow (case 1 and case 2 in Fig.~\ref{fig:growth_flow_density}). For the $n=36$ mode, the normalized growth rate reduces from $0.3313$ to $0.3183$, by nearly $3.92\%$, due to the increase of edge density level from $1.0 \times 10^{19}m^{-3}$ and $4.0 \times 10^{19}m^{-3}$. Next, we consider the flow effect alone for the fixed density profile (case 1 and case 3 in Fig.~\ref{fig:growth_flow_density}). When a toroidal flow specified with $M=0.2$ and $S=30$ is introduced, the normalized growth rate of the $n=36$ mode is found to decrease from $0.3313$ to $0.2745$ by $17.15\%$. Finally, when both the density increase and the toroidal flow mentioned earlier are taken into account for comparison (case 1 and case 4 in Fig.~\ref{fig:growth_flow_density}), the normalized growth rate of the same $n=36$ mode reduces from $0.3313$ to $0.2411$ by $27.2\%$. In another word, the combination of flow and density provides stronger ($27.2\%$) reduction in growth rate than the simple addition of the two individual reductions alone (which would be $3.92\%+17.15\%=21.07\%$). This indicates that the enhanced flow stabilizing effects on ELMs due to increased edge density involves a novel and intrinsic coupling mechanism of these two seemingly unrelated elements.


In summary, we reveal for the first time a new direct role of edge plasma density in the physics of edge localized modes. We find that increasing plasma density can significantly enhance the toroidal flow stabilization of edge localized modes, based on the thorough analysis of a model H-mode equilibrium using NIMROD. The stabilizing effect of toroidal flow alone on high-$n$ modes is shown only effective for flow with large amplitude and shear, far beyond those achievable for modern tokamak such as ITER. However, when plasma density is increased, significant more effective stabilization of high-$n$ ELMs is achieved for all given flow profiles. Such an enhanced flow stabilization of ELMs due to increased plasma density remains effective in full extended MHD model when two fluid effects are included, and when the density profile is non-uniform with edge pedestal. Our finding may resolve the mystery why the NBIs, which bring in both enhanced toroidal rotation and plasma density, are known to have significant effects on ELMs in experiments~\citep{Oyama10,Burrell09}, even though the effects from the toroidal rotation alone are rather weak in theory. Furthermore, our finding may also explain why the ELM mitigation and suppression by toroidal flow appears to be more favorably observed in higher collisionality regime in EAST~\citep{Kong}. The newly found effect of plasma density on edge localized modes may also pose an additional constraint on the choice of optimal density level for the desired edge plasma conditions.

In the analysis reported in this paper, the correction of flow to the force-blanced MHD equilibrium has not been taking into account. It turns out that when the correction of flow to equilibrium is included, our findings on the enhanced flow stabilization of ELMs due to increased plasma density remain valid. The details of that study is being reported elsewhere~\citep{Skcheng17}. In addition, increasing edge plasma density can increase the plasma collisionality and hence change the bootstrap current. This may bring in additional reduction in growth rate of edge localized modes. In this work, we isolate and focus on the effect of the plasma density without potential complications from the additional effects from the current modification. Our work complements the study in~\citep{Xu.Ma.Li14} which examines these effects all together and the work in~\cite{P.Zhu12} that considers only the effect of varying current profile. In future we plan on including the current effects in addition to the density effects on the flow stabilization of ELMs.


\begin{acknowledgments}
The research was supported by the National Magnetic Confinement Fusion Program of China under Grant Nos. 2014GB124002, 2015GB101004, and the 100 Talent Program
of the Chinese Academy of Sciences (CAS). Author D. B. is partially supported by CAS President International Fellowship Initiative (PIFI). Author P. Zhu also acknowledges the supports from U.S. DOE grant Nos. DE-FG02-86ER53218 and DE-FC02-
08ER54975. We are grateful for the support from the NIMROD team. The numerical
calculations in this paper have been done on the super computing system in the
Supercomputing Center of University of Science and Technology of China. This research used resources of the National Energy Research Scientific Computing Center, a DOE Office of Science User Facility supported by the Office of Science of the U.S. Department of Energy under Contract No. DE-AC02-05CH11231.
\end{acknowledgments}

\newpage

\newpage
\renewcommand{\emph}[1]{{\it{#1}}}
\bibliographystyle{apsrev4-1}
\bibliography{cheng}

\begin{thebibliography}{50}%
\makeatletter
\providecommand \@ifxundefined [1]{%
 \@ifx{#1\undefined}
}%
\providecommand \@ifnum [1]{%
 \ifnum #1\expandafter \@firstoftwo
 \else \expandafter \@secondoftwo
 \fi
}%
\providecommand \@ifx [1]{%
 \ifx #1\expandafter \@firstoftwo
 \else \expandafter \@secondoftwo
 \fi
}%
\providecommand \natexlab [1]{#1}%
\providecommand \enquote  [1]{``#1''}%
\providecommand \bibnamefont  [1]{#1}%
\providecommand \bibfnamefont [1]{#1}%
\providecommand \citenamefont [1]{#1}%
\providecommand \href@noop [0]{\@secondoftwo}%
\providecommand \href [0]{\begingroup \@sanitize@url \@href}%
\providecommand \@href[1]{\@@startlink{#1}\@@href}%
\providecommand \@@href[1]{\endgroup#1\@@endlink}%
\providecommand \@sanitize@url [0]{\catcode `\\12\catcode `\$12\catcode
  `\&12\catcode `\#12\catcode `\^12\catcode `\_12\catcode `\%12\relax}%
\providecommand \@@startlink[1]{}%
\providecommand \@@endlink[0]{}%
\providecommand \url  [0]{\begingroup\@sanitize@url \@url }%
\providecommand \@url [1]{\endgroup\@href {#1}{\urlprefix }}%
\providecommand \urlprefix  [0]{URL }%
\providecommand \Eprint [0]{\href }%
\providecommand \doibase [0]{http://dx.doi.org/}%
\providecommand \selectlanguage [0]{\@gobble}%
\providecommand \bibinfo  [0]{\@secondoftwo}%
\providecommand \bibfield  [0]{\@secondoftwo}%
\providecommand \translation [1]{[#1]}%
\providecommand \BibitemOpen [0]{}%
\providecommand \bibitemStop [0]{}%
\providecommand \bibitemNoStop [0]{.\EOS\space}%
\providecommand \EOS [0]{\spacefactor3000\relax}%
\providecommand \BibitemShut  [1]{\csname bibitem#1\endcsname}%
\let\auto@bib@innerbib\@empty
\bibitem [{\citenamefont {Carey}\ and\ \citenamefont
  {Sovinec}(2009)}]{carey09a}%
  \BibitemOpen
  \bibfield  {author} {\bibinfo {author} {\bibfnamefont {C.~S.}\ \bibnamefont
  {Carey}}\ and\ \bibinfo {author} {\bibfnamefont {C.~R.}\ \bibnamefont
  {Sovinec}},\ }\href {\doibase 10.1088/0004-637X/699/1/362} {\bibfield
  {journal} {\bibinfo  {journal} {J. Geophys. Res. Space Physics}\ }\textbf
  {\bibinfo {volume} {699}},\ \bibinfo {pages} {362} (\bibinfo {year}
  {2009})}\BibitemShut {NoStop}%
\bibitem [{\citenamefont {Kunz}\ \emph {et~al.}(2016)\citenamefont {Kunz},
  \citenamefont {Stone},\ and\ \citenamefont {Quataert}}]{Matthew16}%
  \BibitemOpen
  \bibfield  {author} {\bibinfo {author} {\bibfnamefont {M.~W.}\ \bibnamefont
  {Kunz}}, \bibinfo {author} {\bibfnamefont {J.~M.}\ \bibnamefont {Stone}}, \
  and\ \bibinfo {author} {\bibfnamefont {E.}~\bibnamefont {Quataert}},\
  }\href@noop {} {\bibfield  {journal} {\bibinfo  {journal} {Phys. Rev. Lett.}\
  }\textbf {\bibinfo {volume} {117}},\ \bibinfo {pages} {235101} (\bibinfo
  {year} {2016})}\BibitemShut {NoStop}%
\bibitem [{\citenamefont {Angelopoulos}\ \emph {et~al.}(1992)\citenamefont
  {Angelopoulos}, \citenamefont {Baumjohann}, \citenamefont {Kennel},
  \citenamefont {Coroniti}, \citenamefont {Kivelson}, \citenamefont {Pellat},
  \citenamefont {Walker}, \citenamefont {L{\"{u}}hr},\ and\ \citenamefont
  {Paschmann}}]{angelopoulos92a}%
  \BibitemOpen
  \bibfield  {author} {\bibinfo {author} {\bibfnamefont {V.}~\bibnamefont
  {Angelopoulos}}, \bibinfo {author} {\bibfnamefont {W.}~\bibnamefont
  {Baumjohann}}, \bibinfo {author} {\bibfnamefont {C.~F.}\ \bibnamefont
  {Kennel}}, \bibinfo {author} {\bibfnamefont {F.~V.}\ \bibnamefont
  {Coroniti}}, \bibinfo {author} {\bibfnamefont {M.~G.}\ \bibnamefont
  {Kivelson}}, \bibinfo {author} {\bibfnamefont {R.}~\bibnamefont {Pellat}},
  \bibinfo {author} {\bibfnamefont {R.~J.}\ \bibnamefont {Walker}}, \bibinfo
  {author} {\bibfnamefont {H.}~\bibnamefont {L{\"{u}}hr}}, \ and\ \bibinfo
  {author} {\bibfnamefont {G.}~\bibnamefont {Paschmann}},\ }\href@noop {}
  {\bibfield  {journal} {\bibinfo  {journal} {J. Geophys. Res.}\ }\textbf
  {\bibinfo {volume} {97}},\ \bibinfo {pages} {4027} (\bibinfo {year}
  {1992})}\BibitemShut {NoStop}%
\bibitem [{\citenamefont {Angelopoulos}\ \emph {et~al.}(2008)\citenamefont
  {Angelopoulos}, \citenamefont {McFadden}, \citenamefont {Larson},
  \citenamefont {Carlson}, \citenamefont {Mende}, \citenamefont {Frey},
  \citenamefont {Phan}, \citenamefont {Sibeck}, \citenamefont {Glassmeier},
  \citenamefont {Auster}, \citenamefont {Donovan}, \citenamefont {Mann},
  \citenamefont {Rae}, \citenamefont {Russell}, \citenamefont {Runov},
  \citenamefont {Xhou},\ and\ \citenamefont {Kepko}}]{angelopoulos08c}%
  \BibitemOpen
  \bibfield  {author} {\bibinfo {author} {\bibfnamefont {V.}~\bibnamefont
  {Angelopoulos}}, \bibinfo {author} {\bibfnamefont {J.}~\bibnamefont
  {McFadden}}, \bibinfo {author} {\bibfnamefont {D.}~\bibnamefont {Larson}},
  \bibinfo {author} {\bibfnamefont {C.}~\bibnamefont {Carlson}}, \bibinfo
  {author} {\bibfnamefont {S.}~\bibnamefont {Mende}}, \bibinfo {author}
  {\bibfnamefont {H.}~\bibnamefont {Frey}}, \bibinfo {author} {\bibfnamefont
  {T.}~\bibnamefont {Phan}}, \bibinfo {author} {\bibfnamefont {D.}~\bibnamefont
  {Sibeck}}, \bibinfo {author} {\bibfnamefont {K.-H.}\ \bibnamefont
  {Glassmeier}}, \bibinfo {author} {\bibfnamefont {U.}~\bibnamefont {Auster}},
  \bibinfo {author} {\bibfnamefont {E.}~\bibnamefont {Donovan}}, \bibinfo
  {author} {\bibfnamefont {I.}~\bibnamefont {Mann}}, \bibinfo {author}
  {\bibfnamefont {I.}~\bibnamefont {Rae}}, \bibinfo {author} {\bibfnamefont
  {C.}~\bibnamefont {Russell}}, \bibinfo {author} {\bibfnamefont
  {A.}~\bibnamefont {Runov}}, \bibinfo {author} {\bibfnamefont
  {X.}~\bibnamefont {Xhou}}, \ and\ \bibinfo {author} {\bibfnamefont
  {L.}~\bibnamefont {Kepko}},\ }\href {\doibase 10.1126/science.1160495}
  {\bibfield  {journal} {\bibinfo  {journal} {Science}\ }\textbf {\bibinfo
  {volume} {321}},\ \bibinfo {pages} {931} (\bibinfo {year}
  {2008})}\BibitemShut {NoStop}%
\bibitem [{\citenamefont {Nishimura}\ \emph {et~al.}(2016)\citenamefont
  {Nishimura}, \citenamefont {Lyons}, \citenamefont {Nicolls}, \citenamefont
  {Hampton}, \citenamefont {Michell}, \citenamefont {Samara}, \citenamefont
  {Bristow}, \citenamefont {Donovan}, \citenamefont {Spanswick}, \citenamefont
  {Angelopoulos},\ and\ \citenamefont {Mende}}]{nishimuray14b}%
  \BibitemOpen
  \bibfield  {author} {\bibinfo {author} {\bibfnamefont {Y.}~\bibnamefont
  {Nishimura}}, \bibinfo {author} {\bibfnamefont {L.~R.}\ \bibnamefont
  {Lyons}}, \bibinfo {author} {\bibfnamefont {M.~J.}\ \bibnamefont {Nicolls}},
  \bibinfo {author} {\bibfnamefont {D.~L.}\ \bibnamefont {Hampton}}, \bibinfo
  {author} {\bibfnamefont {R.~G.}\ \bibnamefont {Michell}}, \bibinfo {author}
  {\bibfnamefont {M.}~\bibnamefont {Samara}}, \bibinfo {author} {\bibfnamefont
  {W.~A.}\ \bibnamefont {Bristow}}, \bibinfo {author} {\bibfnamefont {E.~F.}\
  \bibnamefont {Donovan}}, \bibinfo {author} {\bibfnamefont {E.}~\bibnamefont
  {Spanswick}}, \bibinfo {author} {\bibfnamefont {V.}~\bibnamefont
  {Angelopoulos}}, \ and\ \bibinfo {author} {\bibfnamefont {S.~B.}\
  \bibnamefont {Mende}},\ }\href {\doibase 10.1002/2014JA019773} {\ \textbf
  {\bibinfo {volume} {119}},\ \bibinfo {pages} {3333} (\bibinfo {year}
  {2016})}\BibitemShut {NoStop}%
\bibitem [{\citenamefont {Nishimura}\ and\ \citenamefont
  {Lyons}(2016)}]{nishimuray16a}%
  \BibitemOpen
  \bibfield  {author} {\bibinfo {author} {\bibfnamefont {Y.}~\bibnamefont
  {Nishimura}}\ and\ \bibinfo {author} {\bibfnamefont {L.~R.}\ \bibnamefont
  {Lyons}},\ }\href {\doibase 10.1002/2015JA022128} {\bibfield  {journal}
  {\bibinfo  {journal} {J. Geophys. Res. Space Physics}\ }\textbf {\bibinfo
  {volume} {121}},\ \bibinfo {pages} {1327} (\bibinfo {year}
  {2016})}\BibitemShut {NoStop}%
\bibitem [{\citenamefont {Fox}\ \emph {et~al.}(2011)\citenamefont {Fox},
  \citenamefont {Bhattacharjee},\ and\ \citenamefont {Germaschewski}}]{Fox11}%
  \BibitemOpen
  \bibfield  {author} {\bibinfo {author} {\bibfnamefont {W.}~\bibnamefont
  {Fox}}, \bibinfo {author} {\bibfnamefont {A.}~\bibnamefont {Bhattacharjee}},
  \ and\ \bibinfo {author} {\bibfnamefont {K.}~\bibnamefont {Germaschewski}},\
  }\href@noop {} {\bibfield  {journal} {\bibinfo  {journal} {Phys. Rev. Lett.}\
  }\textbf {\bibinfo {volume} {106}},\ \bibinfo {pages} {215003} (\bibinfo
  {year} {2011})}\BibitemShut {NoStop}%
\bibitem [{\citenamefont {Plechaty}\ \emph {et~al.}(2013)\citenamefont
  {Plechaty}, \citenamefont {Presura},\ and\ \citenamefont
  {Esaulov}}]{Plechaty13}%
  \BibitemOpen
  \bibfield  {author} {\bibinfo {author} {\bibfnamefont {C.}~\bibnamefont
  {Plechaty}}, \bibinfo {author} {\bibfnamefont {R.}~\bibnamefont {Presura}}, \
  and\ \bibinfo {author} {\bibfnamefont {A.~A.}\ \bibnamefont {Esaulov}},\
  }\href@noop {} {\bibfield  {journal} {\bibinfo  {journal} {Phys. Rev. Lett.}\
  }\textbf {\bibinfo {volume} {111}},\ \bibinfo {pages} {185002} (\bibinfo
  {year} {2013})}\BibitemShut {NoStop}%
\bibitem [{\citenamefont {Zohm}(1996)}]{Zohm96}%
  \BibitemOpen
  \bibfield  {author} {\bibinfo {author} {\bibfnamefont {H.}~\bibnamefont
  {Zohm}},\ }\href@noop {} {\bibfield  {journal} {\bibinfo  {journal} {Plasma
  Physics and Controlled Fusion}\ }\textbf {\bibinfo {volume} {38}},\ \bibinfo
  {pages} {105} (\bibinfo {year} {1996})}\BibitemShut {NoStop}%
\bibitem [{\citenamefont {Burrell}\ \emph {et~al.}(2016)\citenamefont
  {Burrell}, \citenamefont {Barada}, \citenamefont {Chen}, \citenamefont
  {Garofalo}, \citenamefont {Groebner}, \citenamefont {Muscatello},
  \citenamefont {Osborne}, \citenamefont {Petty}, \citenamefont {Rhodes},
  \citenamefont {Snyder}, \citenamefont {Solomon}, \citenamefont {Yan},\ and\
  \citenamefont {Zeng}}]{Burrell16}%
  \BibitemOpen
  \bibfield  {author} {\bibinfo {author} {\bibfnamefont {K.~H.}\ \bibnamefont
  {Burrell}}, \bibinfo {author} {\bibfnamefont {K.}~\bibnamefont {Barada}},
  \bibinfo {author} {\bibfnamefont {X.}~\bibnamefont {Chen}}, \bibinfo {author}
  {\bibfnamefont {A.~M.}\ \bibnamefont {Garofalo}}, \bibinfo {author}
  {\bibfnamefont {R.~J.}\ \bibnamefont {Groebner}}, \bibinfo {author}
  {\bibfnamefont {C.~M.}\ \bibnamefont {Muscatello}}, \bibinfo {author}
  {\bibfnamefont {T.~H.}\ \bibnamefont {Osborne}}, \bibinfo {author}
  {\bibfnamefont {C.~C.}\ \bibnamefont {Petty}}, \bibinfo {author}
  {\bibfnamefont {T.~L.}\ \bibnamefont {Rhodes}}, \bibinfo {author}
  {\bibfnamefont {P.~B.}\ \bibnamefont {Snyder}}, \bibinfo {author}
  {\bibfnamefont {W.~M.}\ \bibnamefont {Solomon}}, \bibinfo {author}
  {\bibfnamefont {Z.}~\bibnamefont {Yan}}, \ and\ \bibinfo {author}
  {\bibfnamefont {L.}~\bibnamefont {Zeng}},\ }\href@noop {} {\bibfield
  {journal} {\bibinfo  {journal} {Physics of Plasmas}\ }\textbf {\bibinfo
  {volume} {23}},\ \bibinfo {pages} {056103} (\bibinfo {year}
  {2016})}\BibitemShut {NoStop}%
\bibitem [{\citenamefont {Terry}(2000)}]{terry00a}%
  \BibitemOpen
  \bibfield  {author} {\bibinfo {author} {\bibfnamefont {P.~W.}\ \bibnamefont
  {Terry}},\ }\href@noop {} {\bibfield  {journal} {\bibinfo  {journal} {Reviews
  of Modern Physics}\ }\textbf {\bibinfo {volume} {72}},\ \bibinfo {pages}
  {109} (\bibinfo {year} {2000})}\BibitemShut {NoStop}%
\bibitem [{\citenamefont {Wagner}\ \emph {et~al.}(1982)\citenamefont {Wagner},
  \citenamefont {Becker}, \citenamefont {Behringer}, \citenamefont {Campbell},
  \citenamefont {Eberhagen}, \citenamefont {Engelhardt}, \citenamefont
  {Fussmann}, \citenamefont {Gehre}, \citenamefont {Gernhardt}, \citenamefont
  {Gierke}, \citenamefont {Haas}, \citenamefont {Huang}, \citenamefont
  {Karger}, \citenamefont {Keilhacker}, \citenamefont {Kl\"uber}, \citenamefont
  {Kornherr}, \citenamefont {Lackner}, \citenamefont {Lisitano}, \citenamefont
  {Lister}, \citenamefont {Mayer}, \citenamefont {Meisel}, \citenamefont
  {M\"uller}, \citenamefont {Murmann}, \citenamefont {Niedermeyer},
  \citenamefont {Poschenrieder}, \citenamefont {Rapp}, \citenamefont {R\"ohr},
  \citenamefont {Schneider}, \citenamefont {Siller}, \citenamefont {Speth},
  \citenamefont {St\"abler}, \citenamefont {Steuer}, \citenamefont {Venus},
  \citenamefont {Vollmer},\ and\ \citenamefont {Y\"u}}]{Wagner82}%
  \BibitemOpen
  \bibfield  {author} {\bibinfo {author} {\bibfnamefont {F.}~\bibnamefont
  {Wagner}}, \bibinfo {author} {\bibfnamefont {G.}~\bibnamefont {Becker}},
  \bibinfo {author} {\bibfnamefont {K.}~\bibnamefont {Behringer}}, \bibinfo
  {author} {\bibfnamefont {D.}~\bibnamefont {Campbell}}, \bibinfo {author}
  {\bibfnamefont {A.}~\bibnamefont {Eberhagen}}, \bibinfo {author}
  {\bibfnamefont {W.}~\bibnamefont {Engelhardt}}, \bibinfo {author}
  {\bibfnamefont {G.}~\bibnamefont {Fussmann}}, \bibinfo {author}
  {\bibfnamefont {O.}~\bibnamefont {Gehre}}, \bibinfo {author} {\bibfnamefont
  {J.}~\bibnamefont {Gernhardt}}, \bibinfo {author} {\bibfnamefont {G.~v.}\
  \bibnamefont {Gierke}}, \bibinfo {author} {\bibfnamefont {G.}~\bibnamefont
  {Haas}}, \bibinfo {author} {\bibfnamefont {M.}~\bibnamefont {Huang}},
  \bibinfo {author} {\bibfnamefont {F.}~\bibnamefont {Karger}}, \bibinfo
  {author} {\bibfnamefont {M.}~\bibnamefont {Keilhacker}}, \bibinfo {author}
  {\bibfnamefont {O.}~\bibnamefont {Kl\"uber}}, \bibinfo {author}
  {\bibfnamefont {M.}~\bibnamefont {Kornherr}}, \bibinfo {author}
  {\bibfnamefont {K.}~\bibnamefont {Lackner}}, \bibinfo {author} {\bibfnamefont
  {G.}~\bibnamefont {Lisitano}}, \bibinfo {author} {\bibfnamefont {G.~G.}\
  \bibnamefont {Lister}}, \bibinfo {author} {\bibfnamefont {H.~M.}\
  \bibnamefont {Mayer}}, \bibinfo {author} {\bibfnamefont {D.}~\bibnamefont
  {Meisel}}, \bibinfo {author} {\bibfnamefont {E.~R.}\ \bibnamefont
  {M\"uller}}, \bibinfo {author} {\bibfnamefont {H.}~\bibnamefont {Murmann}},
  \bibinfo {author} {\bibfnamefont {H.}~\bibnamefont {Niedermeyer}}, \bibinfo
  {author} {\bibfnamefont {W.}~\bibnamefont {Poschenrieder}}, \bibinfo {author}
  {\bibfnamefont {H.}~\bibnamefont {Rapp}}, \bibinfo {author} {\bibfnamefont
  {H.}~\bibnamefont {R\"ohr}}, \bibinfo {author} {\bibfnamefont
  {F.}~\bibnamefont {Schneider}}, \bibinfo {author} {\bibfnamefont
  {G.}~\bibnamefont {Siller}}, \bibinfo {author} {\bibfnamefont
  {E.}~\bibnamefont {Speth}}, \bibinfo {author} {\bibfnamefont
  {A.}~\bibnamefont {St\"abler}}, \bibinfo {author} {\bibfnamefont {K.~H.}\
  \bibnamefont {Steuer}}, \bibinfo {author} {\bibfnamefont {G.}~\bibnamefont
  {Venus}}, \bibinfo {author} {\bibfnamefont {O.}~\bibnamefont {Vollmer}}, \
  and\ \bibinfo {author} {\bibfnamefont {Z.}~\bibnamefont {Y\"u}},\ }\href@noop
  {} {\bibfield  {journal} {\bibinfo  {journal} {Phys. Rev. Lett.}\ }\textbf
  {\bibinfo {volume} {49}},\ \bibinfo {pages} {1408} (\bibinfo {year}
  {1982})}\BibitemShut {NoStop}%
\bibitem [{\citenamefont {Burrell}\ \emph {et~al.}(2005)\citenamefont
  {Burrell}, \citenamefont {Evans}, \citenamefont {Doyle}, \citenamefont
  {Fenstermacher}, \citenamefont {Groebner}, \citenamefont {Leonard},
  \citenamefont {Moyer}, \citenamefont {Osborne}, \citenamefont {Schaffer},
  \citenamefont {Snyder}, \citenamefont {Thomas}, \citenamefont {West},
  \citenamefont {Boedo}, \citenamefont {Garofalo}, \citenamefont {Gohil},
  \citenamefont {Jackson}, \citenamefont {Haye}, \citenamefont {Lasnier},
  \citenamefont {Reimerdes}, \citenamefont {Rhodes}, \citenamefont {Scoville},
  \citenamefont {Solomon}, \citenamefont {Thomas}, \citenamefont {Wang},
  \citenamefont {Watkins},\ and\ \citenamefont {Zeng}}]{Burrell05}%
  \BibitemOpen
  \bibfield  {author} {\bibinfo {author} {\bibfnamefont {K.~H.}\ \bibnamefont
  {Burrell}}, \bibinfo {author} {\bibfnamefont {T.~E.}\ \bibnamefont {Evans}},
  \bibinfo {author} {\bibfnamefont {E.~J.}\ \bibnamefont {Doyle}}, \bibinfo
  {author} {\bibfnamefont {M.~E.}\ \bibnamefont {Fenstermacher}}, \bibinfo
  {author} {\bibfnamefont {R.~J.}\ \bibnamefont {Groebner}}, \bibinfo {author}
  {\bibfnamefont {A.~W.}\ \bibnamefont {Leonard}}, \bibinfo {author}
  {\bibfnamefont {R.~A.}\ \bibnamefont {Moyer}}, \bibinfo {author}
  {\bibfnamefont {T.~H.}\ \bibnamefont {Osborne}}, \bibinfo {author}
  {\bibfnamefont {M.~J.}\ \bibnamefont {Schaffer}}, \bibinfo {author}
  {\bibfnamefont {P.~B.}\ \bibnamefont {Snyder}}, \bibinfo {author}
  {\bibfnamefont {P.~R.}\ \bibnamefont {Thomas}}, \bibinfo {author}
  {\bibfnamefont {W.~P.}\ \bibnamefont {West}}, \bibinfo {author}
  {\bibfnamefont {J.~A.}\ \bibnamefont {Boedo}}, \bibinfo {author}
  {\bibfnamefont {A.~M.}\ \bibnamefont {Garofalo}}, \bibinfo {author}
  {\bibfnamefont {P.}~\bibnamefont {Gohil}}, \bibinfo {author} {\bibfnamefont
  {G.~L.}\ \bibnamefont {Jackson}}, \bibinfo {author} {\bibfnamefont
  {R.~J.~L.}\ \bibnamefont {Haye}}, \bibinfo {author} {\bibfnamefont {C.~J.}\
  \bibnamefont {Lasnier}}, \bibinfo {author} {\bibfnamefont {H.}~\bibnamefont
  {Reimerdes}}, \bibinfo {author} {\bibfnamefont {T.~L.}\ \bibnamefont
  {Rhodes}}, \bibinfo {author} {\bibfnamefont {J.~T.}\ \bibnamefont
  {Scoville}}, \bibinfo {author} {\bibfnamefont {W.~M.}\ \bibnamefont
  {Solomon}}, \bibinfo {author} {\bibfnamefont {D.~M.}\ \bibnamefont {Thomas}},
  \bibinfo {author} {\bibfnamefont {G.}~\bibnamefont {Wang}}, \bibinfo {author}
  {\bibfnamefont {J.~G.}\ \bibnamefont {Watkins}}, \ and\ \bibinfo {author}
  {\bibfnamefont {L.}~\bibnamefont {Zeng}},\ }\href@noop {} {\bibfield
  {journal} {\bibinfo  {journal} {Plasma Physics and Controlled Fusion}\
  }\textbf {\bibinfo {volume} {47}},\ \bibinfo {pages} {B37} (\bibinfo {year}
  {2005})}\BibitemShut {NoStop}%
\bibitem [{\citenamefont {Burrell}\ \emph {et~al.}(2009)\citenamefont
  {Burrell}, \citenamefont {Osborne}, \citenamefont {Snyder}, \citenamefont
  {West}, \citenamefont {Fenstermacher}, \citenamefont {Groebner},
  \citenamefont {Gohil}, \citenamefont {Leonard},\ and\ \citenamefont
  {Solomon}}]{Burrell09}%
  \BibitemOpen
  \bibfield  {author} {\bibinfo {author} {\bibfnamefont {K.}~\bibnamefont
  {Burrell}}, \bibinfo {author} {\bibfnamefont {T.}~\bibnamefont {Osborne}},
  \bibinfo {author} {\bibfnamefont {P.}~\bibnamefont {Snyder}}, \bibinfo
  {author} {\bibfnamefont {W.}~\bibnamefont {West}}, \bibinfo {author}
  {\bibfnamefont {M.}~\bibnamefont {Fenstermacher}}, \bibinfo {author}
  {\bibfnamefont {R.}~\bibnamefont {Groebner}}, \bibinfo {author}
  {\bibfnamefont {P.}~\bibnamefont {Gohil}}, \bibinfo {author} {\bibfnamefont
  {A.}~\bibnamefont {Leonard}}, \ and\ \bibinfo {author} {\bibfnamefont
  {W.}~\bibnamefont {Solomon}},\ }\href@noop {} {\bibfield  {journal} {\bibinfo
   {journal} {Nuclear Fusion}\ }\textbf {\bibinfo {volume} {49}},\ \bibinfo
  {pages} {085024} (\bibinfo {year} {2009})}\BibitemShut {NoStop}%
\bibitem [{\citenamefont {Leonard}(2014)}]{Leonard14}%
  \BibitemOpen
  \bibfield  {author} {\bibinfo {author} {\bibfnamefont {A.~W.}\ \bibnamefont
  {Leonard}},\ }\href@noop {} {\bibfield  {journal} {\bibinfo  {journal}
  {Physics of Plasmas}\ }\textbf {\bibinfo {volume} {21}},\ \bibinfo {eid}
  {090501} (\bibinfo {year} {2014})}\BibitemShut {NoStop}%
\bibitem [{\citenamefont {Loarte}\ \emph {et~al.}(2007)\citenamefont {Loarte},
  \citenamefont {Lipschultz}, \citenamefont {Kukushkin}, \citenamefont
  {Matthews}, \citenamefont {Stangeby}, \citenamefont {Asakura}, \citenamefont
  {Counsell}, \citenamefont {Federici}, \citenamefont {Kallenbach},
  \citenamefont {Krieger}, \citenamefont {Mahdavi}, \citenamefont {Philipps},
  \citenamefont {Reiter}, \citenamefont {Roth}, \citenamefont {Strachan},
  \citenamefont {Whyte}, \citenamefont {Doerner}, \citenamefont {Eich},
  \citenamefont {Fundamenski}, \citenamefont {Herrmann}, \citenamefont
  {Fenstermacher}, \citenamefont {Ghendrih}, \citenamefont {Groth},
  \citenamefont {Kirschner}, \citenamefont {Konoshima}, \citenamefont
  {LaBombard}, \citenamefont {Lang}, \citenamefont {Leonard}, \citenamefont
  {Monier-Garbet}, \citenamefont {Neu}, \citenamefont {Pacher}, \citenamefont
  {Pegourie}, \citenamefont {Pitts}, \citenamefont {Takamura}, \citenamefont
  {Terry}, \citenamefont {Tsitrone}, \citenamefont {the ITPA Scrape-off
  Layer},\ and\ \citenamefont {Group}}]{Loarte07}%
  \BibitemOpen
  \bibfield  {author} {\bibinfo {author} {\bibfnamefont {A.}~\bibnamefont
  {Loarte}}, \bibinfo {author} {\bibfnamefont {B.}~\bibnamefont {Lipschultz}},
  \bibinfo {author} {\bibfnamefont {A.}~\bibnamefont {Kukushkin}}, \bibinfo
  {author} {\bibfnamefont {G.}~\bibnamefont {Matthews}}, \bibinfo {author}
  {\bibfnamefont {P.}~\bibnamefont {Stangeby}}, \bibinfo {author}
  {\bibfnamefont {N.}~\bibnamefont {Asakura}}, \bibinfo {author} {\bibfnamefont
  {G.}~\bibnamefont {Counsell}}, \bibinfo {author} {\bibfnamefont
  {G.}~\bibnamefont {Federici}}, \bibinfo {author} {\bibfnamefont
  {A.}~\bibnamefont {Kallenbach}}, \bibinfo {author} {\bibfnamefont
  {K.}~\bibnamefont {Krieger}}, \bibinfo {author} {\bibfnamefont
  {A.}~\bibnamefont {Mahdavi}}, \bibinfo {author} {\bibfnamefont
  {V.}~\bibnamefont {Philipps}}, \bibinfo {author} {\bibfnamefont
  {D.}~\bibnamefont {Reiter}}, \bibinfo {author} {\bibfnamefont
  {J.}~\bibnamefont {Roth}}, \bibinfo {author} {\bibfnamefont {J.}~\bibnamefont
  {Strachan}}, \bibinfo {author} {\bibfnamefont {D.}~\bibnamefont {Whyte}},
  \bibinfo {author} {\bibfnamefont {R.}~\bibnamefont {Doerner}}, \bibinfo
  {author} {\bibfnamefont {T.}~\bibnamefont {Eich}}, \bibinfo {author}
  {\bibfnamefont {W.}~\bibnamefont {Fundamenski}}, \bibinfo {author}
  {\bibfnamefont {A.}~\bibnamefont {Herrmann}}, \bibinfo {author}
  {\bibfnamefont {M.}~\bibnamefont {Fenstermacher}}, \bibinfo {author}
  {\bibfnamefont {P.}~\bibnamefont {Ghendrih}}, \bibinfo {author}
  {\bibfnamefont {M.}~\bibnamefont {Groth}}, \bibinfo {author} {\bibfnamefont
  {A.}~\bibnamefont {Kirschner}}, \bibinfo {author} {\bibfnamefont
  {S.}~\bibnamefont {Konoshima}}, \bibinfo {author} {\bibfnamefont
  {B.}~\bibnamefont {LaBombard}}, \bibinfo {author} {\bibfnamefont
  {P.}~\bibnamefont {Lang}}, \bibinfo {author} {\bibfnamefont {A.}~\bibnamefont
  {Leonard}}, \bibinfo {author} {\bibfnamefont {P.}~\bibnamefont
  {Monier-Garbet}}, \bibinfo {author} {\bibfnamefont {R.}~\bibnamefont {Neu}},
  \bibinfo {author} {\bibfnamefont {H.}~\bibnamefont {Pacher}}, \bibinfo
  {author} {\bibfnamefont {B.}~\bibnamefont {Pegourie}}, \bibinfo {author}
  {\bibfnamefont {R.}~\bibnamefont {Pitts}}, \bibinfo {author} {\bibfnamefont
  {S.}~\bibnamefont {Takamura}}, \bibinfo {author} {\bibfnamefont
  {J.}~\bibnamefont {Terry}}, \bibinfo {author} {\bibfnamefont
  {E.}~\bibnamefont {Tsitrone}}, \bibinfo {author} {\bibnamefont {the ITPA
  Scrape-off Layer}}, \ and\ \bibinfo {author} {\bibfnamefont {D.~P.~T.}\
  \bibnamefont {Group}},\ }\href@noop {} {\bibfield  {journal} {\bibinfo
  {journal} {Nuclear Fusion}\ }\textbf {\bibinfo {volume} {47}},\ \bibinfo
  {pages} {S203} (\bibinfo {year} {2007})}\BibitemShut {NoStop}%
\bibitem [{\citenamefont {Oyama}\ \emph {et~al.}(2005)\citenamefont {Oyama},
  \citenamefont {Sakamoto}, \citenamefont {Isayama}, \citenamefont {Takechi},
  \citenamefont {Gohil}, \citenamefont {Lao}, \citenamefont {Snyder},
  \citenamefont {Fujita}, \citenamefont {Ide}, \citenamefont {Kamada},
  \citenamefont {Miura}, \citenamefont {Oikawa}, \citenamefont {Suzuki},
  \citenamefont {Takenaga}, \citenamefont {Toi},\ and\ \citenamefont {the
  JT-60~Team}}]{Oyama05}%
  \BibitemOpen
  \bibfield  {author} {\bibinfo {author} {\bibfnamefont {N.}~\bibnamefont
  {Oyama}}, \bibinfo {author} {\bibfnamefont {Y.}~\bibnamefont {Sakamoto}},
  \bibinfo {author} {\bibfnamefont {A.}~\bibnamefont {Isayama}}, \bibinfo
  {author} {\bibfnamefont {M.}~\bibnamefont {Takechi}}, \bibinfo {author}
  {\bibfnamefont {P.}~\bibnamefont {Gohil}}, \bibinfo {author} {\bibfnamefont
  {L.}~\bibnamefont {Lao}}, \bibinfo {author} {\bibfnamefont {P.}~\bibnamefont
  {Snyder}}, \bibinfo {author} {\bibfnamefont {T.}~\bibnamefont {Fujita}},
  \bibinfo {author} {\bibfnamefont {S.}~\bibnamefont {Ide}}, \bibinfo {author}
  {\bibfnamefont {Y.}~\bibnamefont {Kamada}}, \bibinfo {author} {\bibfnamefont
  {Y.}~\bibnamefont {Miura}}, \bibinfo {author} {\bibfnamefont
  {T.}~\bibnamefont {Oikawa}}, \bibinfo {author} {\bibfnamefont
  {T.}~\bibnamefont {Suzuki}}, \bibinfo {author} {\bibfnamefont
  {H.}~\bibnamefont {Takenaga}}, \bibinfo {author} {\bibfnamefont
  {K.}~\bibnamefont {Toi}}, \ and\ \bibinfo {author} {\bibnamefont {the
  JT-60~Team}},\ }\href@noop {} {\bibfield  {journal} {\bibinfo  {journal}
  {Nuclear Fusion}\ }\textbf {\bibinfo {volume} {45}},\ \bibinfo {pages} {871}
  (\bibinfo {year} {2005})}\BibitemShut {NoStop}%
\bibitem [{\citenamefont {Cooper}(1988)}]{Cooper88}%
  \BibitemOpen
  \bibfield  {author} {\bibinfo {author} {\bibfnamefont {W.~A.}\ \bibnamefont
  {Cooper}},\ }\href@noop {} {\bibfield  {journal} {\bibinfo  {journal} {Plasma
  Physics and Controlled Fusion}\ }\textbf {\bibinfo {volume} {30}},\ \bibinfo
  {pages} {1805} (\bibinfo {year} {1988})}\BibitemShut {NoStop}%
\bibitem [{\citenamefont {Waelbroeck}\ and\ \citenamefont
  {Chen}(1991)}]{Wealbroeck91}%
  \BibitemOpen
  \bibfield  {author} {\bibinfo {author} {\bibfnamefont {F.~L.}\ \bibnamefont
  {Waelbroeck}}\ and\ \bibinfo {author} {\bibfnamefont {L.}~\bibnamefont
  {Chen}},\ }\href@noop {} {\bibfield  {journal} {\bibinfo  {journal} {Physics
  of Fluids B}\ }\textbf {\bibinfo {volume} {3}},\ \bibinfo {pages} {601}
  (\bibinfo {year} {1991})}\BibitemShut {NoStop}%
\bibitem [{\citenamefont {Furukawa}\ and\ \citenamefont
  {Tokuda}(2005)}]{Furukawa05}%
  \BibitemOpen
  \bibfield  {author} {\bibinfo {author} {\bibfnamefont {M.}~\bibnamefont
  {Furukawa}}\ and\ \bibinfo {author} {\bibfnamefont {S.}~\bibnamefont
  {Tokuda}},\ }\href@noop {} {\bibfield  {journal} {\bibinfo  {journal} {Phys.
  Rev. Lett.}\ }\textbf {\bibinfo {volume} {94}},\ \bibinfo {pages} {175001}
  (\bibinfo {year} {2005})}\BibitemShut {NoStop}%
\bibitem [{\citenamefont {Aiba}\ \emph {et~al.}(2009)\citenamefont {Aiba},
  \citenamefont {Tokuda}, \citenamefont {Furukawa}, \citenamefont {Oyama},\
  and\ \citenamefont {Ozeki}}]{Aiba09}%
  \BibitemOpen
  \bibfield  {author} {\bibinfo {author} {\bibfnamefont {N.}~\bibnamefont
  {Aiba}}, \bibinfo {author} {\bibfnamefont {S.}~\bibnamefont {Tokuda}},
  \bibinfo {author} {\bibfnamefont {M.}~\bibnamefont {Furukawa}}, \bibinfo
  {author} {\bibfnamefont {N.}~\bibnamefont {Oyama}}, \ and\ \bibinfo {author}
  {\bibfnamefont {T.}~\bibnamefont {Ozeki}},\ }\href@noop {} {\bibfield
  {journal} {\bibinfo  {journal} {Nuclear Fusion}\ }\textbf {\bibinfo {volume}
  {49}},\ \bibinfo {pages} {065015} (\bibinfo {year} {2009})}\BibitemShut
  {NoStop}%
\bibitem [{\citenamefont {Aiba}\ \emph {et~al.}(2010)\citenamefont {Aiba},
  \citenamefont {Furukawa}, \citenamefont {Hirota},\ and\ \citenamefont
  {Tokuda}}]{Aiba10}%
  \BibitemOpen
  \bibfield  {author} {\bibinfo {author} {\bibfnamefont {N.}~\bibnamefont
  {Aiba}}, \bibinfo {author} {\bibfnamefont {M.}~\bibnamefont {Furukawa}},
  \bibinfo {author} {\bibfnamefont {M.}~\bibnamefont {Hirota}}, \ and\ \bibinfo
  {author} {\bibfnamefont {S.}~\bibnamefont {Tokuda}},\ }\href@noop {}
  {\bibfield  {journal} {\bibinfo  {journal} {Nuclear Fusion}\ }\textbf
  {\bibinfo {volume} {50}},\ \bibinfo {pages} {045002} (\bibinfo {year}
  {2010})}\BibitemShut {NoStop}%
\bibitem [{\citenamefont {Aiba}\ \emph {et~al.}(2011)\citenamefont {Aiba},
  \citenamefont {Furukawa}, \citenamefont {Hirota}, \citenamefont {Oyama},
  \citenamefont {Kojima}, \citenamefont {Tokuda},\ and\ \citenamefont
  {Yagi}}]{Aiba11}%
  \BibitemOpen
  \bibfield  {author} {\bibinfo {author} {\bibfnamefont {N.}~\bibnamefont
  {Aiba}}, \bibinfo {author} {\bibfnamefont {M.}~\bibnamefont {Furukawa}},
  \bibinfo {author} {\bibfnamefont {M.}~\bibnamefont {Hirota}}, \bibinfo
  {author} {\bibfnamefont {N.}~\bibnamefont {Oyama}}, \bibinfo {author}
  {\bibfnamefont {A.}~\bibnamefont {Kojima}}, \bibinfo {author} {\bibfnamefont
  {S.}~\bibnamefont {Tokuda}}, \ and\ \bibinfo {author} {\bibfnamefont
  {M.}~\bibnamefont {Yagi}},\ }\href@noop {} {\bibfield  {journal} {\bibinfo
  {journal} {Nuclear Fusion}\ }\textbf {\bibinfo {volume} {51}},\ \bibinfo
  {pages} {073012} (\bibinfo {year} {2011})}\BibitemShut {NoStop}%
\bibitem [{\citenamefont {Xi}\ \emph {et~al.}(2012)\citenamefont {Xi},
  \citenamefont {Xu}, \citenamefont {Wang},\ and\ \citenamefont {Xia}}]{Xi12}%
  \BibitemOpen
  \bibfield  {author} {\bibinfo {author} {\bibfnamefont {P.~W.}\ \bibnamefont
  {Xi}}, \bibinfo {author} {\bibfnamefont {X.~Q.}\ \bibnamefont {Xu}}, \bibinfo
  {author} {\bibfnamefont {X.~G.}\ \bibnamefont {Wang}}, \ and\ \bibinfo
  {author} {\bibfnamefont {T.~Y.}\ \bibnamefont {Xia}},\ }\href@noop {}
  {\bibfield  {journal} {\bibinfo  {journal} {Physics of Plasmas}\ }\textbf
  {\bibinfo {volume} {19}},\ \bibinfo {eid} {092503} (\bibinfo {year}
  {2012})}\BibitemShut {NoStop}%
\bibitem [{\citenamefont {Xia}\ and\ \citenamefont {Xu}(2013)}]{Xia13}%
  \BibitemOpen
  \bibfield  {author} {\bibinfo {author} {\bibfnamefont {T.~Y.}\ \bibnamefont
  {Xia}}\ and\ \bibinfo {author} {\bibfnamefont {X.~Q.}\ \bibnamefont {Xu}},\
  }\href@noop {} {\bibfield  {journal} {\bibinfo  {journal} {Physics of
  Plasmas}\ }\textbf {\bibinfo {volume} {20}},\ \bibinfo {eid} {052102}
  (\bibinfo {year} {2013})}\BibitemShut {NoStop}%
\bibitem [{\citenamefont {Mazzucato}\ \emph {et~al.}(1996)\citenamefont
  {Mazzucato}, \citenamefont {Batha}, \citenamefont {Beer}, \citenamefont
  {Bell}, \citenamefont {Bell}, \citenamefont {Budny}, \citenamefont {Bush},
  \citenamefont {Hahm}, \citenamefont {Hammett}, \citenamefont {Levinton},
  \citenamefont {Nazikian}, \citenamefont {Park}, \citenamefont {Rewoldt},
  \citenamefont {Schmidt}, \citenamefont {Synakowski}, \citenamefont {Tang},
  \citenamefont {Taylor},\ and\ \citenamefont {Zarnstorff}}]{mazzucato93a}%
  \BibitemOpen
  \bibfield  {author} {\bibinfo {author} {\bibfnamefont {E.}~\bibnamefont
  {Mazzucato}}, \bibinfo {author} {\bibfnamefont {S.}~\bibnamefont {Batha}},
  \bibinfo {author} {\bibfnamefont {M.}~\bibnamefont {Beer}}, \bibinfo {author}
  {\bibfnamefont {M.}~\bibnamefont {Bell}}, \bibinfo {author} {\bibfnamefont
  {R.}~\bibnamefont {Bell}}, \bibinfo {author} {\bibfnamefont {R.}~\bibnamefont
  {Budny}}, \bibinfo {author} {\bibfnamefont {C.}~\bibnamefont {Bush}},
  \bibinfo {author} {\bibfnamefont {T.}~\bibnamefont {Hahm}}, \bibinfo {author}
  {\bibfnamefont {G.}~\bibnamefont {Hammett}}, \bibinfo {author} {\bibfnamefont
  {F.}~\bibnamefont {Levinton}}, \bibinfo {author} {\bibfnamefont
  {R.}~\bibnamefont {Nazikian}}, \bibinfo {author} {\bibfnamefont
  {H.}~\bibnamefont {Park}}, \bibinfo {author} {\bibfnamefont {G.}~\bibnamefont
  {Rewoldt}}, \bibinfo {author} {\bibfnamefont {G.}~\bibnamefont {Schmidt}},
  \bibinfo {author} {\bibfnamefont {E.}~\bibnamefont {Synakowski}}, \bibinfo
  {author} {\bibfnamefont {W.}~\bibnamefont {Tang}}, \bibinfo {author}
  {\bibfnamefont {G.}~\bibnamefont {Taylor}}, \ and\ \bibinfo {author}
  {\bibfnamefont {M.}~\bibnamefont {Zarnstorff}},\ }\href@noop {} {\bibfield
  {journal} {\bibinfo  {journal} {Physical Review Letters}\ }\textbf {\bibinfo
  {volume} {77}},\ \bibinfo {pages} {3145} (\bibinfo {year}
  {1996})}\BibitemShut {NoStop}%
\bibitem [{\citenamefont {Rice}\ \emph {et~al.}(1996)\citenamefont {Rice},
  \citenamefont {Lazarus}, \citenamefont {Austin}, \citenamefont {Burrell},
  \citenamefont {Casper}, \citenamefont {Groebner}, \citenamefont {Gohil},
  \citenamefont {Forest}, \citenamefont {Ikezi}, \citenamefont {Lao},
  \citenamefont {Mauel}, \citenamefont {Navratil}, \citenamefont {Stallard},
  \citenamefont {Strait},\ and\ \citenamefont {Taylor}}]{rice96a}%
  \BibitemOpen
  \bibfield  {author} {\bibinfo {author} {\bibfnamefont {B.~W.}\ \bibnamefont
  {Rice}}, \bibinfo {author} {\bibfnamefont {E.~A.}\ \bibnamefont {Lazarus}},
  \bibinfo {author} {\bibfnamefont {M.~E.}\ \bibnamefont {Austin}}, \bibinfo
  {author} {\bibfnamefont {K.~H.}\ \bibnamefont {Burrell}}, \bibinfo {author}
  {\bibfnamefont {T.~A.}\ \bibnamefont {Casper}}, \bibinfo {author}
  {\bibfnamefont {R.~J.}\ \bibnamefont {Groebner}}, \bibinfo {author}
  {\bibfnamefont {P.}~\bibnamefont {Gohil}}, \bibinfo {author} {\bibfnamefont
  {C.~B.}\ \bibnamefont {Forest}}, \bibinfo {author} {\bibfnamefont
  {H.}~\bibnamefont {Ikezi}}, \bibinfo {author} {\bibfnamefont {L.~L.}\
  \bibnamefont {Lao}}, \bibinfo {author} {\bibfnamefont {M.~E.}\ \bibnamefont
  {Mauel}}, \bibinfo {author} {\bibfnamefont {G.~A.}\ \bibnamefont {Navratil}},
  \bibinfo {author} {\bibfnamefont {B.~W.}\ \bibnamefont {Stallard}}, \bibinfo
  {author} {\bibfnamefont {E.~J.}\ \bibnamefont {Strait}}, \ and\ \bibinfo
  {author} {\bibfnamefont {T.~S.}\ \bibnamefont {Taylor}},\ }\href@noop {}
  {\bibfield  {journal} {\bibinfo  {journal} {Nuclear Fusion}\ }\textbf
  {\bibinfo {volume} {36}},\ \bibinfo {pages} {1271} (\bibinfo {year}
  {1996})}\BibitemShut {NoStop}%
\bibitem [{\citenamefont {Gormezano}\ and\ \citenamefont {the
  JET~Team}(1997)}]{gormezano96a}%
  \BibitemOpen
  \bibfield  {author} {\bibinfo {author} {\bibfnamefont {C.}~\bibnamefont
  {Gormezano}}\ and\ \bibinfo {author} {\bibnamefont {the JET~Team}},\ }in\
  \href@noop {} {\emph {\bibinfo {booktitle} {Proceedings of the Sixteenth IAEA
  Fusion Energy Conference, (Montr\'{e}al, Canada, 1996)}}},\ Vol.~\bibinfo
  {volume} {1}\ (\bibinfo  {publisher} {IAEA},\ \bibinfo {address} {Vienna},\
  \bibinfo {year} {1997})\ p.\ \bibinfo {pages} {487},\ \bibinfo {note} {paper
  IAEA-CN-64/A5-5}\BibitemShut {NoStop}%
\bibitem [{\citenamefont {Conway}\ \emph {et~al.}(2000)\citenamefont {Conway},
  \citenamefont {Borba}, \citenamefont {Alper}, \citenamefont {Bartlett},
  \citenamefont {Gormezano}, \citenamefont {von Hellermann}, \citenamefont
  {Maas}, \citenamefont {Parail}, \citenamefont {Smeulders},\ and\
  \citenamefont {Zastrow}}]{conway00a}%
  \BibitemOpen
  \bibfield  {author} {\bibinfo {author} {\bibfnamefont {G.~D.}\ \bibnamefont
  {Conway}}, \bibinfo {author} {\bibfnamefont {D.~N.}\ \bibnamefont {Borba}},
  \bibinfo {author} {\bibfnamefont {B.}~\bibnamefont {Alper}}, \bibinfo
  {author} {\bibfnamefont {D.~V.}\ \bibnamefont {Bartlett}}, \bibinfo {author}
  {\bibfnamefont {C.}~\bibnamefont {Gormezano}}, \bibinfo {author}
  {\bibfnamefont {M.~G.}\ \bibnamefont {von Hellermann}}, \bibinfo {author}
  {\bibfnamefont {A.~C.}\ \bibnamefont {Maas}}, \bibinfo {author}
  {\bibfnamefont {V.~V.}\ \bibnamefont {Parail}}, \bibinfo {author}
  {\bibfnamefont {P.}~\bibnamefont {Smeulders}}, \ and\ \bibinfo {author}
  {\bibfnamefont {K.-D.}\ \bibnamefont {Zastrow}},\ }\href@noop {} {\bibfield
  {journal} {\bibinfo  {journal} {Physical Review Letters}\ }\textbf {\bibinfo
  {volume} {84}},\ \bibinfo {pages} {1463} (\bibinfo {year}
  {2000})}\BibitemShut {NoStop}%
\bibitem [{\citenamefont {de~Vries}\ \emph {et~al.}(2006)\citenamefont
  {de~Vries}, \citenamefont {Rantamäki}, \citenamefont {Giroud}, \citenamefont
  {Asp}, \citenamefont {Corrigan}, \citenamefont {Eriksson}, \citenamefont
  {de~Greef}, \citenamefont {Jenkins}, \citenamefont {Knoops}, \citenamefont
  {Mantica}, \citenamefont {Nordman}, \citenamefont {Strand}, \citenamefont
  {Tala}, \citenamefont {Weiland}, \citenamefont {Zastrow},\ and\ \citenamefont
  {Contributors}}]{Vries06}%
  \BibitemOpen
  \bibfield  {author} {\bibinfo {author} {\bibfnamefont {P.~C.}\ \bibnamefont
  {de~Vries}}, \bibinfo {author} {\bibfnamefont {K.~M.}\ \bibnamefont
  {Rantamäki}}, \bibinfo {author} {\bibfnamefont {C.}~\bibnamefont {Giroud}},
  \bibinfo {author} {\bibfnamefont {E.}~\bibnamefont {Asp}}, \bibinfo {author}
  {\bibfnamefont {G.}~\bibnamefont {Corrigan}}, \bibinfo {author}
  {\bibfnamefont {A.}~\bibnamefont {Eriksson}}, \bibinfo {author}
  {\bibfnamefont {M.}~\bibnamefont {de~Greef}}, \bibinfo {author}
  {\bibfnamefont {I.}~\bibnamefont {Jenkins}}, \bibinfo {author} {\bibfnamefont
  {H.~C.~M.}\ \bibnamefont {Knoops}}, \bibinfo {author} {\bibfnamefont
  {P.}~\bibnamefont {Mantica}}, \bibinfo {author} {\bibfnamefont
  {H.}~\bibnamefont {Nordman}}, \bibinfo {author} {\bibfnamefont
  {P.}~\bibnamefont {Strand}}, \bibinfo {author} {\bibfnamefont
  {T.}~\bibnamefont {Tala}}, \bibinfo {author} {\bibfnamefont {J.}~\bibnamefont
  {Weiland}}, \bibinfo {author} {\bibfnamefont {K.-D.}\ \bibnamefont
  {Zastrow}}, \ and\ \bibinfo {author} {\bibfnamefont {J.~E.}\ \bibnamefont
  {Contributors}},\ }\href@noop {} {\bibfield  {journal} {\bibinfo  {journal}
  {Plasma Physics and Controlled Fusion}\ }\textbf {\bibinfo {volume} {48}},\
  \bibinfo {pages} {1693} (\bibinfo {year} {2006})}\BibitemShut {NoStop}%
\bibitem [{\citenamefont {Rice}\ \emph {et~al.}(2007)\citenamefont {Rice},
  \citenamefont {Ince-Cushman}, \citenamefont {deGrassie}, \citenamefont
  {Eriksson}, \citenamefont {Sakamoto}, \citenamefont {Scarabosio},
  \citenamefont {Bortolon}, \citenamefont {Burrell}, \citenamefont {Duval},
  \citenamefont {Fenzi-Bonizec}, \citenamefont {Greenwald}, \citenamefont
  {Groebner}, \citenamefont {Hoang}, \citenamefont {Koide}, \citenamefont
  {Marmar}, \citenamefont {Pochelon},\ and\ \citenamefont {Podpaly}}]{Rice07}%
  \BibitemOpen
  \bibfield  {author} {\bibinfo {author} {\bibfnamefont {J.}~\bibnamefont
  {Rice}}, \bibinfo {author} {\bibfnamefont {A.}~\bibnamefont {Ince-Cushman}},
  \bibinfo {author} {\bibfnamefont {J.}~\bibnamefont {deGrassie}}, \bibinfo
  {author} {\bibfnamefont {L.-G.}\ \bibnamefont {Eriksson}}, \bibinfo {author}
  {\bibfnamefont {Y.}~\bibnamefont {Sakamoto}}, \bibinfo {author}
  {\bibfnamefont {A.}~\bibnamefont {Scarabosio}}, \bibinfo {author}
  {\bibfnamefont {A.}~\bibnamefont {Bortolon}}, \bibinfo {author}
  {\bibfnamefont {K.}~\bibnamefont {Burrell}}, \bibinfo {author} {\bibfnamefont
  {B.}~\bibnamefont {Duval}}, \bibinfo {author} {\bibfnamefont
  {C.}~\bibnamefont {Fenzi-Bonizec}}, \bibinfo {author} {\bibfnamefont
  {M.}~\bibnamefont {Greenwald}}, \bibinfo {author} {\bibfnamefont
  {R.}~\bibnamefont {Groebner}}, \bibinfo {author} {\bibfnamefont
  {G.}~\bibnamefont {Hoang}}, \bibinfo {author} {\bibfnamefont
  {Y.}~\bibnamefont {Koide}}, \bibinfo {author} {\bibfnamefont
  {E.}~\bibnamefont {Marmar}}, \bibinfo {author} {\bibfnamefont
  {A.}~\bibnamefont {Pochelon}}, \ and\ \bibinfo {author} {\bibfnamefont
  {Y.}~\bibnamefont {Podpaly}},\ }\href@noop {} {\bibfield  {journal} {\bibinfo
   {journal} {Nuclear Fusion}\ }\textbf {\bibinfo {volume} {47}},\ \bibinfo
  {pages} {1618} (\bibinfo {year} {2007})}\BibitemShut {NoStop}%
\bibitem [{\citenamefont {Politzer}\ \emph {et~al.}(2008)\citenamefont
  {Politzer}, \citenamefont {Petty}, \citenamefont {Jayakumar}, \citenamefont
  {Luce}, \citenamefont {Wade}, \citenamefont {DeBoo}, \citenamefont {Ferron},
  \citenamefont {Gohil}, \citenamefont {Holcomb}, \citenamefont {Hyatt},
  \citenamefont {Kinsey}, \citenamefont {Haye}, \citenamefont {Makowski},\ and\
  \citenamefont {Petrie}}]{Politzer08}%
  \BibitemOpen
  \bibfield  {author} {\bibinfo {author} {\bibfnamefont {P.}~\bibnamefont
  {Politzer}}, \bibinfo {author} {\bibfnamefont {C.}~\bibnamefont {Petty}},
  \bibinfo {author} {\bibfnamefont {R.}~\bibnamefont {Jayakumar}}, \bibinfo
  {author} {\bibfnamefont {T.}~\bibnamefont {Luce}}, \bibinfo {author}
  {\bibfnamefont {M.}~\bibnamefont {Wade}}, \bibinfo {author} {\bibfnamefont
  {J.}~\bibnamefont {DeBoo}}, \bibinfo {author} {\bibfnamefont
  {J.}~\bibnamefont {Ferron}}, \bibinfo {author} {\bibfnamefont
  {P.}~\bibnamefont {Gohil}}, \bibinfo {author} {\bibfnamefont
  {C.}~\bibnamefont {Holcomb}}, \bibinfo {author} {\bibfnamefont
  {A.}~\bibnamefont {Hyatt}}, \bibinfo {author} {\bibfnamefont
  {J.}~\bibnamefont {Kinsey}}, \bibinfo {author} {\bibfnamefont {R.~L.}\
  \bibnamefont {Haye}}, \bibinfo {author} {\bibfnamefont {M.}~\bibnamefont
  {Makowski}}, \ and\ \bibinfo {author} {\bibfnamefont {T.}~\bibnamefont
  {Petrie}},\ }\href@noop {} {\bibfield  {journal} {\bibinfo  {journal}
  {Nuclear Fusion}\ }\textbf {\bibinfo {volume} {48}},\ \bibinfo {pages}
  {075001} (\bibinfo {year} {2008})}\BibitemShut {NoStop}%
\bibitem [{\citenamefont {Wan}\ \emph {et~al.}(2014)\citenamefont {Wan},
  \citenamefont {Ding}, \citenamefont {Qian}, \citenamefont {Li}, \citenamefont
  {Xiao},\ and\ \citenamefont {Xu}}]{Wan14}%
  \BibitemOpen
  \bibfield  {author} {\bibinfo {author} {\bibfnamefont {B.}~\bibnamefont
  {Wan}}, \bibinfo {author} {\bibfnamefont {S.}~\bibnamefont {Ding}}, \bibinfo
  {author} {\bibfnamefont {J.}~\bibnamefont {Qian}}, \bibinfo {author}
  {\bibfnamefont {G.}~\bibnamefont {Li}}, \bibinfo {author} {\bibfnamefont
  {B.}~\bibnamefont {Xiao}}, \ and\ \bibinfo {author} {\bibfnamefont
  {G.}~\bibnamefont {Xu}},\ }\href {\doibase 10.1109/TPS.2013.2296939}
  {\bibfield  {journal} {\bibinfo  {journal} {IEEE Transactions on Plasma
  Science}\ }\textbf {\bibinfo {volume} {42}},\ \bibinfo {pages} {495}
  (\bibinfo {year} {2014})}\BibitemShut {NoStop}%
\bibitem [{\citenamefont {Chan}\ \emph {et~al.}(2015)\citenamefont {Chan},
  \citenamefont {Costley}, \citenamefont {Wan}, \citenamefont {Garofalo},\ and\
  \citenamefont {Leuer}}]{Chan15}%
  \BibitemOpen
  \bibfield  {author} {\bibinfo {author} {\bibfnamefont {V.}~\bibnamefont
  {Chan}}, \bibinfo {author} {\bibfnamefont {A.}~\bibnamefont {Costley}},
  \bibinfo {author} {\bibfnamefont {B.}~\bibnamefont {Wan}}, \bibinfo {author}
  {\bibfnamefont {A.}~\bibnamefont {Garofalo}}, \ and\ \bibinfo {author}
  {\bibfnamefont {J.}~\bibnamefont {Leuer}},\ }\href
  {http://stacks.iop.org/0029-5515/55/i=2/a=023017} {\bibfield  {journal}
  {\bibinfo  {journal} {Nuclear Fusion}\ }\textbf {\bibinfo {volume} {55}},\
  \bibinfo {pages} {023017} (\bibinfo {year} {2015})}\BibitemShut {NoStop}%
\bibitem [{\citenamefont {Oyama}\ \emph {et~al.}(2010)\citenamefont {Oyama},
  \citenamefont {Kojima}, \citenamefont {Aiba}, \citenamefont {Horton},
  \citenamefont {Isayama}, \citenamefont {Kamiya}, \citenamefont {Urano},
  \citenamefont {Sakamoto}, \citenamefont {Kamada},\ and\ \citenamefont {the
  JT-60~Team}}]{Oyama10}%
  \BibitemOpen
  \bibfield  {author} {\bibinfo {author} {\bibfnamefont {N.}~\bibnamefont
  {Oyama}}, \bibinfo {author} {\bibfnamefont {A.}~\bibnamefont {Kojima}},
  \bibinfo {author} {\bibfnamefont {N.}~\bibnamefont {Aiba}}, \bibinfo {author}
  {\bibfnamefont {L.}~\bibnamefont {Horton}}, \bibinfo {author} {\bibfnamefont
  {A.}~\bibnamefont {Isayama}}, \bibinfo {author} {\bibfnamefont
  {K.}~\bibnamefont {Kamiya}}, \bibinfo {author} {\bibfnamefont
  {H.}~\bibnamefont {Urano}}, \bibinfo {author} {\bibfnamefont
  {Y.}~\bibnamefont {Sakamoto}}, \bibinfo {author} {\bibfnamefont
  {Y.}~\bibnamefont {Kamada}}, \ and\ \bibinfo {author} {\bibnamefont {the
  JT-60~Team}},\ }\href@noop {} {\bibfield  {journal} {\bibinfo  {journal}
  {Nuclear Fusion}\ }\textbf {\bibinfo {volume} {50}},\ \bibinfo {pages}
  {064014} (\bibinfo {year} {2010})}\BibitemShut {NoStop}%
\bibitem [{\citenamefont {Kong}\ \emph {et~al.}(2016)\citenamefont {Kong},
  \citenamefont {Xu}, \citenamefont {Huang}, \citenamefont {Gao},\ and\
  \citenamefont {Chen}}]{Kong}%
  \BibitemOpen
  \bibfield  {author} {\bibinfo {author} {\bibfnamefont {D.~F.}\ \bibnamefont
  {Kong}}, \bibinfo {author} {\bibfnamefont {X.~Q.}\ \bibnamefont {Xu}},
  \bibinfo {author} {\bibfnamefont {C.~B.}\ \bibnamefont {Huang}}, \bibinfo
  {author} {\bibfnamefont {X.}~\bibnamefont {Gao}}, \ and\ \bibinfo {author}
  {\bibfnamefont {J.~G.}\ \bibnamefont {Chen}},\ }in\ \href@noop {} {\emph
  {\bibinfo {booktitle} {Proceedings of the 43th EPS Conference on Plasma
  Physics, Leuven, Belgium, July 4-8, 2016}}}\ (\bibinfo {year}
  {2016})\BibitemShut {NoStop}%
\bibitem [{\citenamefont {Sovinec}\ \emph {et~al.}(2004)\citenamefont
  {Sovinec}, \citenamefont {Glasser}, \citenamefont {Gianakon}, \citenamefont
  {Barnes}, \citenamefont {Nebel}, \citenamefont {Kruger}, \citenamefont
  {Schnack}, \citenamefont {Plimpton}, \citenamefont {Tarditi},\ and\
  \citenamefont {Chu}}]{Sovinec04}%
  \BibitemOpen
  \bibfield  {author} {\bibinfo {author} {\bibfnamefont {C.}~\bibnamefont
  {Sovinec}}, \bibinfo {author} {\bibfnamefont {A.}~\bibnamefont {Glasser}},
  \bibinfo {author} {\bibfnamefont {T.}~\bibnamefont {Gianakon}}, \bibinfo
  {author} {\bibfnamefont {D.}~\bibnamefont {Barnes}}, \bibinfo {author}
  {\bibfnamefont {R.}~\bibnamefont {Nebel}}, \bibinfo {author} {\bibfnamefont
  {S.}~\bibnamefont {Kruger}}, \bibinfo {author} {\bibfnamefont
  {D.}~\bibnamefont {Schnack}}, \bibinfo {author} {\bibfnamefont
  {S.}~\bibnamefont {Plimpton}}, \bibinfo {author} {\bibfnamefont
  {A.}~\bibnamefont {Tarditi}}, \ and\ \bibinfo {author} {\bibfnamefont
  {M.}~\bibnamefont {Chu}},\ }\href@noop {} {\bibfield  {journal} {\bibinfo
  {journal} {Journal of Computational Physics}\ }\textbf {\bibinfo {volume}
  {195}},\ \bibinfo {pages} {355 } (\bibinfo {year} {2004})}\BibitemShut
  {NoStop}%
\bibitem [{\citenamefont {Sovinec}\ and\ \citenamefont
  {King}(2010)}]{Sovinec10}%
  \BibitemOpen
  \bibfield  {author} {\bibinfo {author} {\bibfnamefont {C.}~\bibnamefont
  {Sovinec}}\ and\ \bibinfo {author} {\bibfnamefont {J.}~\bibnamefont {King}},\
  }\href@noop {} {\bibfield  {journal} {\bibinfo  {journal} {Journal of
  Computational Physics}\ }\textbf {\bibinfo {volume} {229}},\ \bibinfo {pages}
  {5803 } (\bibinfo {year} {2010})}\BibitemShut {NoStop}%
\bibitem [{\citenamefont {Brennan}\ \emph {et~al.}(2006)\citenamefont
  {Brennan}, \citenamefont {Kruger}, \citenamefont {Schnack}, \citenamefont
  {Sovinec},\ and\ \citenamefont {Pankin}}]{Brennan06}%
  \BibitemOpen
  \bibfield  {author} {\bibinfo {author} {\bibfnamefont {D.~P.}\ \bibnamefont
  {Brennan}}, \bibinfo {author} {\bibfnamefont {S.~E.}\ \bibnamefont {Kruger}},
  \bibinfo {author} {\bibfnamefont {D.~D.}\ \bibnamefont {Schnack}}, \bibinfo
  {author} {\bibfnamefont {C.~R.}\ \bibnamefont {Sovinec}}, \ and\ \bibinfo
  {author} {\bibfnamefont {A.}~\bibnamefont {Pankin}},\ }\href@noop {}
  {\bibfield  {journal} {\bibinfo  {journal} {J. Phys. : Conf. Series}\
  }\textbf {\bibinfo {volume} {46}},\ \bibinfo {pages} {63} (\bibinfo {year}
  {2006})}\BibitemShut {NoStop}%
\bibitem [{\citenamefont {Zhu}\ and\ \citenamefont {Raeder}(2013)}]{Zhu13}%
  \BibitemOpen
  \bibfield  {author} {\bibinfo {author} {\bibfnamefont {P.}~\bibnamefont
  {Zhu}}\ and\ \bibinfo {author} {\bibfnamefont {J.}~\bibnamefont {Raeder}},\
  }\href@noop {} {\bibfield  {journal} {\bibinfo  {journal} {Phys. Rev. Lett.}\
  }\textbf {\bibinfo {volume} {110}},\ \bibinfo {pages} {235005} (\bibinfo
  {year} {2013})}\BibitemShut {NoStop}%
\bibitem [{\citenamefont {Burke}\ \emph {et~al.}(2010)\citenamefont {Burke},
  \citenamefont {Kruger}, \citenamefont {Hegna}, \citenamefont {Zhu},
  \citenamefont {Snyder}, \citenamefont {Sovinec},\ and\ \citenamefont
  {Howell}}]{Burke10}%
  \BibitemOpen
  \bibfield  {author} {\bibinfo {author} {\bibfnamefont {B.~J.}\ \bibnamefont
  {Burke}}, \bibinfo {author} {\bibfnamefont {S.~E.}\ \bibnamefont {Kruger}},
  \bibinfo {author} {\bibfnamefont {C.~C.}\ \bibnamefont {Hegna}}, \bibinfo
  {author} {\bibfnamefont {P.}~\bibnamefont {Zhu}}, \bibinfo {author}
  {\bibfnamefont {P.~B.}\ \bibnamefont {Snyder}}, \bibinfo {author}
  {\bibfnamefont {C.~R.}\ \bibnamefont {Sovinec}}, \ and\ \bibinfo {author}
  {\bibfnamefont {E.~C.}\ \bibnamefont {Howell}},\ }\href@noop {} {\bibfield
  {journal} {\bibinfo  {journal} {Phys. Plasmas}\ }\textbf {\bibinfo {volume}
  {17}},\ \bibinfo {pages} {032103} (\bibinfo {year} {2010})}\BibitemShut
  {NoStop}%
\bibitem [{\citenamefont {Ebrahimi}\ and\ \citenamefont
  {Raman}(2015)}]{Ebrahimi15}%
  \BibitemOpen
  \bibfield  {author} {\bibinfo {author} {\bibfnamefont {F.}~\bibnamefont
  {Ebrahimi}}\ and\ \bibinfo {author} {\bibfnamefont {R.}~\bibnamefont
  {Raman}},\ }\href@noop {} {\bibfield  {journal} {\bibinfo  {journal} {Phys.
  Rev. Lett.}\ }\textbf {\bibinfo {volume} {114}},\ \bibinfo {pages} {205003}
  (\bibinfo {year} {2015})}\BibitemShut {NoStop}%
\bibitem [{\citenamefont {Izzo}\ \emph {et~al.}(2015)\citenamefont {Izzo},
  \citenamefont {Parks}, \citenamefont {Eidietis}, \citenamefont {Shiraki},
  \citenamefont {Hollmann}, \citenamefont {Commaux}, \citenamefont {Granetz},
  \citenamefont {Humphreys}, \citenamefont {Lasnier}, \citenamefont {Moyer},
  \citenamefont {Paz-Soldan}, \citenamefont {Raman},\ and\ \citenamefont
  {Strait}}]{Izzo15}%
  \BibitemOpen
  \bibfield  {author} {\bibinfo {author} {\bibfnamefont {V.}~\bibnamefont
  {Izzo}}, \bibinfo {author} {\bibfnamefont {P.}~\bibnamefont {Parks}},
  \bibinfo {author} {\bibfnamefont {N.}~\bibnamefont {Eidietis}}, \bibinfo
  {author} {\bibfnamefont {D.}~\bibnamefont {Shiraki}}, \bibinfo {author}
  {\bibfnamefont {E.}~\bibnamefont {Hollmann}}, \bibinfo {author}
  {\bibfnamefont {N.}~\bibnamefont {Commaux}}, \bibinfo {author} {\bibfnamefont
  {R.}~\bibnamefont {Granetz}}, \bibinfo {author} {\bibfnamefont
  {D.}~\bibnamefont {Humphreys}}, \bibinfo {author} {\bibfnamefont
  {C.}~\bibnamefont {Lasnier}}, \bibinfo {author} {\bibfnamefont
  {R.}~\bibnamefont {Moyer}}, \bibinfo {author} {\bibfnamefont
  {C.}~\bibnamefont {Paz-Soldan}}, \bibinfo {author} {\bibfnamefont
  {R.}~\bibnamefont {Raman}}, \ and\ \bibinfo {author} {\bibfnamefont
  {E.}~\bibnamefont {Strait}},\ }\href@noop {} {\bibfield  {journal} {\bibinfo
  {journal} {Nucl. Fusion}\ }\textbf {\bibinfo {volume} {55}},\ \bibinfo
  {pages} {073032} (\bibinfo {year} {2015})}\BibitemShut {NoStop}%
\bibitem [{\citenamefont {Zhu}\ \emph {et~al.}(2008)\citenamefont {Zhu},
  \citenamefont {Schnack}, \citenamefont {Ebrahimi}, \citenamefont {Zweibel},
  \citenamefont {Suzuki}, \citenamefont {Hegna},\ and\ \citenamefont
  {Sovinec}}]{Zhu08}%
  \BibitemOpen
  \bibfield  {author} {\bibinfo {author} {\bibfnamefont {P.}~\bibnamefont
  {Zhu}}, \bibinfo {author} {\bibfnamefont {D.~D.}\ \bibnamefont {Schnack}},
  \bibinfo {author} {\bibfnamefont {F.}~\bibnamefont {Ebrahimi}}, \bibinfo
  {author} {\bibfnamefont {E.~G.}\ \bibnamefont {Zweibel}}, \bibinfo {author}
  {\bibfnamefont {M.}~\bibnamefont {Suzuki}}, \bibinfo {author} {\bibfnamefont
  {C.~C.}\ \bibnamefont {Hegna}}, \ and\ \bibinfo {author} {\bibfnamefont
  {C.~R.}\ \bibnamefont {Sovinec}},\ }\href@noop {} {\bibfield  {journal}
  {\bibinfo  {journal} {Phys. Rev. Lett.}\ }\textbf {\bibinfo {volume} {101}},\
  \bibinfo {pages} {085005} (\bibinfo {year} {2008})}\BibitemShut {NoStop}%
\bibitem [{\citenamefont {Miller}\ \emph {et~al.}(1997)\citenamefont {Miller},
  \citenamefont {Lin-Liu}, \citenamefont {Turnbull}, \citenamefont {Chan},
  \citenamefont {Pearlstein}, \citenamefont {Sauter},\ and\ \citenamefont
  {Villard}}]{Miller97}%
  \BibitemOpen
  \bibfield  {author} {\bibinfo {author} {\bibfnamefont {R.~L.}\ \bibnamefont
  {Miller}}, \bibinfo {author} {\bibfnamefont {Y.~R.}\ \bibnamefont {Lin-Liu}},
  \bibinfo {author} {\bibfnamefont {A.~D.}\ \bibnamefont {Turnbull}}, \bibinfo
  {author} {\bibfnamefont {V.~S.}\ \bibnamefont {Chan}}, \bibinfo {author}
  {\bibfnamefont {L.~D.}\ \bibnamefont {Pearlstein}}, \bibinfo {author}
  {\bibfnamefont {O.}~\bibnamefont {Sauter}}, \ and\ \bibinfo {author}
  {\bibfnamefont {L.}~\bibnamefont {Villard}},\ }\href@noop {} {\bibfield
  {journal} {\bibinfo  {journal} {Physics of Plasmas}\ }\textbf {\bibinfo
  {volume} {4}},\ \bibinfo {pages} {1062} (\bibinfo {year} {1997})}\BibitemShut
  {NoStop}%
\bibitem [{\citenamefont {King}\ \emph {et~al.}(2016)\citenamefont {King},
  \citenamefont {Pankin}, \citenamefont {Kruger},\ and\ \citenamefont
  {Snyder}}]{King16}%
  \BibitemOpen
  \bibfield  {author} {\bibinfo {author} {\bibfnamefont {J.~R.}\ \bibnamefont
  {King}}, \bibinfo {author} {\bibfnamefont {A.~Y.}\ \bibnamefont {Pankin}},
  \bibinfo {author} {\bibfnamefont {S.~E.}\ \bibnamefont {Kruger}}, \ and\
  \bibinfo {author} {\bibfnamefont {P.~B.}\ \bibnamefont {Snyder}},\
  }\href@noop {} {\bibfield  {journal} {\bibinfo  {journal} {Physics of
  Plasmas}\ }\textbf {\bibinfo {volume} {23}},\ \bibinfo {eid} {062123}
  (\bibinfo {year} {2016})}\BibitemShut {NoStop}%
\bibitem [{\citenamefont {Xu}\ \emph {et~al.}(2014)\citenamefont {Xu},
  \citenamefont {Ma},\ and\ \citenamefont {Li}}]{Xu.Ma.Li14}%
  \BibitemOpen
  \bibfield  {author} {\bibinfo {author} {\bibfnamefont {X.~Q.}\ \bibnamefont
  {Xu}}, \bibinfo {author} {\bibfnamefont {J.~F.}\ \bibnamefont {Ma}}, \ and\
  \bibinfo {author} {\bibfnamefont {G.~Q.}\ \bibnamefont {Li}},\ }\href@noop {}
  {\bibfield  {journal} {\bibinfo  {journal} {Physics of Plasmas}\ }\textbf
  {\bibinfo {volume} {21}},\ \bibinfo {eid} {120704} (\bibinfo {year}
  {2014})}\BibitemShut {NoStop}%
\bibitem [{\citenamefont {Zweben}\ \emph {et~al.}(2014)\citenamefont {Zweben},
  \citenamefont {Stotler}, \citenamefont {Bell}, \citenamefont {Davis},
  \citenamefont {Kaye}, \citenamefont {LeBlanc}, \citenamefont {Maqueda},
  \citenamefont {Meier}, \citenamefont {Munsat}, \citenamefont {Ren},
  \citenamefont {Sabbagh}, \citenamefont {Sechrest}, \citenamefont {Smith},\
  and\ \citenamefont {Soukhanovskii}}]{Zweben14}%
  \BibitemOpen
  \bibfield  {author} {\bibinfo {author} {\bibfnamefont {S.~J.}\ \bibnamefont
  {Zweben}}, \bibinfo {author} {\bibfnamefont {D.~P.}\ \bibnamefont {Stotler}},
  \bibinfo {author} {\bibfnamefont {R.~E.}\ \bibnamefont {Bell}}, \bibinfo
  {author} {\bibfnamefont {W.~M.}\ \bibnamefont {Davis}}, \bibinfo {author}
  {\bibfnamefont {S.~M.}\ \bibnamefont {Kaye}}, \bibinfo {author}
  {\bibfnamefont {B.~P.}\ \bibnamefont {LeBlanc}}, \bibinfo {author}
  {\bibfnamefont {R.~J.}\ \bibnamefont {Maqueda}}, \bibinfo {author}
  {\bibfnamefont {E.~T.}\ \bibnamefont {Meier}}, \bibinfo {author}
  {\bibfnamefont {T.}~\bibnamefont {Munsat}}, \bibinfo {author} {\bibfnamefont
  {Y.}~\bibnamefont {Ren}}, \bibinfo {author} {\bibfnamefont {S.~A.}\
  \bibnamefont {Sabbagh}}, \bibinfo {author} {\bibfnamefont {Y.}~\bibnamefont
  {Sechrest}}, \bibinfo {author} {\bibfnamefont {D.~R.}\ \bibnamefont {Smith}},
  \ and\ \bibinfo {author} {\bibfnamefont {V.}~\bibnamefont {Soukhanovskii}},\
  }\href@noop {} {\bibfield  {journal} {\bibinfo  {journal} {Plasma Physics and
  Controlled Fusion}\ }\textbf {\bibinfo {volume} {56}},\ \bibinfo {pages}
  {095010} (\bibinfo {year} {2014})}\BibitemShut {NoStop}%
\bibitem [{\citenamefont {Cheng}\ \emph {et~al.}(2017)\citenamefont {Cheng},
  \citenamefont {Zhu}, \citenamefont {Banerjee},\ and\ \citenamefont
  {Li}}]{Skcheng17}%
  \BibitemOpen
  \bibfield  {author} {\bibinfo {author} {\bibfnamefont {S.-K.}\ \bibnamefont
  {Cheng}}, \bibinfo {author} {\bibfnamefont {P.}~\bibnamefont {Zhu}}, \bibinfo
  {author} {\bibfnamefont {D.}~\bibnamefont {Banerjee}}, \ and\ \bibinfo
  {author} {\bibfnamefont {H.}~\bibnamefont {Li}},\ }\href@noop {} {\enquote
  {\bibinfo {title} {Effects of toroidal sheared flow and plasma density on
  edge localized instabilities},}\ } (\bibinfo {year} {2017}),\ \bibinfo {note}
  {submitted to~\pop}\BibitemShut {NoStop}%
\bibitem [{\citenamefont {Zhu}\ \emph {et~al.}(2012)\citenamefont {Zhu},
  \citenamefont {Hegna},\ and\ \citenamefont {Sovinec}}]{P.Zhu12}%
  \BibitemOpen
  \bibfield  {author} {\bibinfo {author} {\bibfnamefont {P.}~\bibnamefont
  {Zhu}}, \bibinfo {author} {\bibfnamefont {C.~C.}\ \bibnamefont {Hegna}}, \
  and\ \bibinfo {author} {\bibfnamefont {C.~R.}\ \bibnamefont {Sovinec}},\
  }\href@noop {} {\bibfield  {journal} {\bibinfo  {journal} {Physics of
  Plasmas}\ }\textbf {\bibinfo {volume} {19}},\ \bibinfo {eid} {032503}
  (\bibinfo {year} {2012})}\BibitemShut {NoStop}%
\end{thebibliography}%
\newpage


\clearpage


\clearpage


\clearpage





\clearpage


\clearpage


\clearpage


\clearpage

\end{document}